\documentclass[10pt, a4paper, twocolumn]{article} 

\usepackage[english]{babel} 
\usepackage[colorlinks,allcolors=blue]{hyperref}
\usepackage{microtype} 

\usepackage{amsmath,amsfonts,amsthm} 

\usepackage[svgnames,table,xcdraw]{xcolor} 

\usepackage[hang, small, labelfont=bf, up, textfont=it]{caption} 

\usepackage{booktabs} 

\usepackage{lastpage} 

\usepackage{graphicx} 
\usepackage{xurl}

\usepackage{enumitem} 
\setlist{noitemsep} 

\usepackage{sectsty} 
\allsectionsfont{\usefont{OT1}{phv}{b}{n}} 

\usepackage{geometry} 

\usepackage[T1]{fontenc} 
\usepackage[utf8]{inputenc} 

\usepackage{XCharter} 

\usepackage{fancyhdr} 
\fancypagestyle{firstpage}{ 
	\fancyhf{}
}

\newcommand{\authorstyle}[1]{{\large\usefont{OT1}{phv}{b}{n}\color{DarkBlue}#1}} 

\newcommand{\institution}[1]{{\footnotesize\usefont{OT1}{phv}{m}{sl}\color{Black}#1}} 

\newcommand{\parz}[1]{\ensuremath{\left(#1\right)}}

\usepackage{titling} 

\newcommand{\HorRule}{\color{DarkGoldenrod}\rule{\linewidth}{1pt}} 

\pretitle{
	\vspace{-30pt} 
	\HorRule\vspace{10pt} 
	\fontsize{28}{36}\usefont{OT1}{phv}{b}{n}\selectfont 
	\color{DarkBlue} 
}

\posttitle{\par\vskip 15pt} 

\preauthor{} 

\postauthor{ 
	\vspace{10pt} 
	\par\HorRule 
	\vspace{20pt} 
}

\usepackage{lettrine} 
\usepackage{fix-cm}	

\newcommand{\initial}[1]{ 
	\lettrine[lines=3,findent=4pt,nindent=0pt]{
		\color{DarkGoldenrod}
		{#1}
	}{}%
}

\usepackage{xstring} 

\newcommand{\lettrineabstract}[1]{
	\StrLeft{#1}{1}[\firstletter] 
	\initial{\firstletter}\textbf{\StrGobbleLeft{#1}{1}} 
}

\newcommand{\bquote}[1]{{\color{DarkBlue}{ \begin{quote} ``#1'' \end{quote}}}}

\usepackage[autostyle=true]{csquotes} 

\title{A self-governing, self-regulating 
system for assessing scientific predictive power} 

\author{
	\authorstyle{Ted C. Rogers\textsuperscript{1}} 
	\newline\newline 
	\textsuperscript{1}\institution{Department of Physics, Old Dominion University, Norfolk, VA 23529, USA}\\ 
	\text{Email: trogers@odu.edu} \\ \text{\href{https://orcid.org/0000-0002-0762-0275}{ORCID: 0000-0002-0762-0275}}
}

\date{\today} 

\begin{document}

\maketitle 

\thispagestyle{firstpage} 


\lettrineabstract{I propose a method for tracking and assessing scientific progress using a prediction consensus algorithm designed for the purpose. The protocol obviates the need for centralized referees to generate scientific questions, gather predictions, and assess the accuracy or success of those predictions. It relies instead on crowd wisdom and a system of checks and balances for all tasks. 
It is intended to take the form of a web-based, searchable database.
I describe a prototype implementation that I call Ex Quaerum. The main purpose of the present document is to motivate the project, to explain its underlying philosophy, to explain the details of the consensus protocol on which it is based, to describe plans for its future development, and ultimately to attract additional collaborators.}

\section{The problem of measuring \\ scientific progress}
\label{s.problem}

\subsection{Science and incentives}

Science in the 21st century is characterized by the following trends:
\begin{itemize}
\item \underline{Increasing complexity}: Learning curves are increasingly steep and it is becoming more difficult for individuals to manage and process new knowledge. As such, information overload is a growing problem~\cite{reisz2022loss,fortunato2018science,jones2009burden}. In the most mature scientific disciplines, it can now take decades for a scientist to accumulate all the knowledge and experience necessary to make well-informed judgments regarding broad trends in their fields. For outsiders and non-experts, such judgments are often impossible. \vspace{.1in}
\item \underline{Increasing specialization}: Partly due to the increased complexity, researchers are driven to focus on much narrower subfields than they have been in the past~\cite{hyper,spec1,bateman2015different}. 
As a result, relevant expertise tends to be fragmented across many separate individuals who may or may not communicate effectively. 
\vspace{.1in}
\item \underline{Cross-disciplinary collaboration}: Progress relies more heavily on synergy between groups from very different scientific fields, often working together within large scientific mega groups. Frequently, scientists from one field or subfield (or subsubfield) must place a high degree of trust in work from fields very different from their own. Breakdowns of communication become more probable in these environments. 
\vspace{.1in}
\item \underline{Growth in both size and cost}: 
A growing fraction of scientists' time and effort is spent finding resources and allocating them. Those resources then need to be distributed across (sometimes collaborating but 
sometimes competing) expert subfields.
\vspace{.1in}
\item \underline{A widening division of labor} exists between the work of scientific theory or hypothesis development and the work of testing, verifying, and/or refuting those theories. Tangled layers of assumptions often separate statements of fundamental ideas and assumptions from concrete tests of them, and the layers tend to grow more numerous, and the tests less direct, as a scientific field matures.
\vspace{.1in}
\item \underline{Increasing professional competition}: Especially in fields with limited professional opportunities, scientists are driven more than ever to engage in promotional activities. 
This leads to inflation in the amount of hype in the scientific literature~\cite{mazin2022inverse,vinkers_use_2015}, which can sometimes evolve into distorted pictures of the science itself.
\end{itemize}

Taken altogether, the above makes assessing genuine overall progress in today’s scientific fields very difficult for decision makers. Traditional figures of merit like citation counts track progress at best indirectly, and they are prone to the distorting influences of large-group social dynamics~\cite{chu2021slowed,higginson2016current,grimes2018modelling,serra2021nonreplicable}. Cognitive biases~\cite{begley_six_2013,nuzzo_how_2015}, group-think, and misaligned incentive structures~\cite{belluz_7_2016,smaldino2016natural,gopalakrishna2022prevalence} have all been documented over the past few decades in commentary about the so-called  \href{https://en.wikipedia.org/wiki/Replication_crisis}{“replication crisis”}~\cite{ioannidis2005most,ritchie_science_2020,allison_reproducibility_2016,amaral_reproducibility_2021,colling_statistical_2021}, and they have been implicated in problems with how science is done more broadly.  \href{https://en.wikipedia.org/wiki/Metascience}{Metascientific} research has also exposed some of the limitations of 
traditional peer review and publication systems as tools for measuring scientific progress and implementing quality control~\cite{fortunato2018science,schooler_metascience_2014,chu_too_2018,nielsen2021global,serra2021nonreplicable,elite,teplitskiy2022status}. 

It is urgent that scientific communities address these problems directly, not only for the sake of practicing scientists but for anyone with an interest in supporting, investing in, or participating in the best possible scientific research. Stakeholders include government and private funding agencies, scientific journals, students deciding whether to embark on a career in a specific scientific topic, or members of the general public concerned about the reliability of public and private scientific institutions. A continued lack of transparency and accountability threatens to undermine confidence in science very broadly. 

This project is meant to be one of hopefully many efforts by practicing scientists to address some of these very general problems. 


\subsection{Help from predictions?}

 It is rarely ever clear what exactly constitutes ``successful'' research, especially when dealing with very complex subjects. Even the role of empirical testability is hotly debated in, for example, some subfields of physics~\cite{dawid_string_2013,baggott_farewell_2013,hossenfelder_lost_2018,false}. In other scientific disciplines, conclusions once thought to be firmly rooted in empirical confirmation, and sometimes even standardized into medical practice, are later found to be false or questionable. Instances of this happen with especially unsettling frequency in the medical sciences~\cite{herrera-perez_meta-research_2019,prasad_decade_2013,harris2017rigor,dozens,errington2021reproducibility}. 
 
 While the problem of replication has been thus far most extensively documented and studied in medicine, the life sciences, and psychology, one can find evidence that similar problems are far more widespread than is yet realized in other fields as well, for example in nuclear and particle physics~\cite{Stone:2000an,Hicks:2012zz,Junk:2020azi,Fowlie:2021aew,Junk2021Response}. 
 In some cases, however, the highly technical nature of a field  creates barriers to anyone attempting to identify, understand, and fix any methodological problems that might exist. It is nonetheless evident that it is not enough to simply assess scientists’ own reported research outcomes in formal publications or reports. There is a need for more direct and transparent indicators of progress from within the science itself.

With the above in mind, it is natural to ask about scientific  predictions, given that a major source of science’s traditional claim to authority is its ability to accurately predict events, be they the outcomes of specific experiments, the course of natural phenomena like epidemics and the climate, or even discoveries of new mathematical theorems. When predictions are consistently refuted, a healthy scientific research community is identifiable by its readiness to update underlying theories and assumptions. 

Not all scientific research is organized around predictions, or course, and exploratory research can be just as valuable as confirmatory research. However, many (or most) large scientific projects today are promoted explicitly on the grounds that they can \emph{test or prove} specific hypotheses and/or theories. And a claim that a research project will test a scientific theory is, at its core, a claim that it will in some way confirm or refute predictions. The power of this kind of confirmatory research lies in the way that it reinforces or challenges prior expectations. It either solidifies existing ideas or creates fertile ground for new ones. Therefore, a natural way to try to track scientific progress is to understand exactly how this process of prediction confirmation/refutation unfolds in practice and how it affects the evolution of scientific ideas.  

But assessing the scientific value of predictions is not simple. For one, simply making arbitrary predictions and determining whether they are confirmed or refuted does not necessarily provide scientifically valuable information. 
To turn a study of scientific predictions into an effective evaluation tool, one must ask at least the following of any claims to predictive power: 
\begin{itemize}
\item What \emph{exactly} was predicted and what was the motivation for predicting it? 
\item How difficult was it to predict? 
\item How was the prediction connected to the theories? 
\item How relevant were they to scientists’ research programs?
\item To what extent did the outcome of the prediction lead scientists to update their assumptions or theoretical beliefs? \item Finally, does the research lead to growing consensus among the scientists?
\end{itemize}

In practice, claims of predictive power are often vague, imprecise, or merely qualitative. Metascientific research has also exposed limitations in experts' own judgments about their own predictions. (See, for example, ~\cite{burgman_trusting_2016,tetlock_superforecasting_2016}). It can be difficult to straightforwardly interpret reported statistics, given the prevalence of phenomena such as \href{https://en.wikipedia.org/wiki/HARKing}{HARKing}, \href{https://en.wikipedia.org/wiki/Postdiction}{postdicting}, \href{https://en.wikipedia.org/wiki/Data_dredging}{p-hacking}, \href{https://en.wikipedia.org/wiki/Publication_bias}{publication bias}, the \href{https://en.wikipedia.org/wiki/Look-elsewhere_effect}{look-elsewhere effect}, and similar problems. 
My proposal, which I will begin to explain in the next section, is my own attempt to contribute to improvement efforts. Its more immediate goal is to improve the usefulness of prediction-tracking as a tool for measuring progress in confirmatory research. 

There are, of course, many existing methods for mitigating problems like those listed above. 
For example, \href{https://en.wikipedia.org/wiki/Preregistration_(science)}{preregistration} of clinical trials is a requirement for much medical research in the United States that receives federal funding. Fields like experimental high energy particle physics use \href{https://en.wikipedia.org/wiki/Blinded_experiment}{blind analysis}~\cite{maccoun_blind_2015} to mitigate \href{https://en.wikipedia.org/wiki/Confirmation_bias}{confirmation bias}. (An interesting implementation of blinding called \href{https://www.ligo.org/news/blind-injection.php}{blind-injection} was used in the LIGO experiment.) Double blind \href{https://en.wikipedia.org/wiki/Placebo}{placebo} controlled studies are the gold standard in much medical research. 

There is also a long history of prediction competitions being used to test expert judgments, and some metascientific projects have used them to study the predictive power of experts in fields like the social sciences. One of the best known is the \href{https://en.wikipedia.org/wiki/The_Good_Judgment_Project}{Good Judgment Open Project} (GJO)~\cite{gjo}, which aggregates numerical predictions mostly regarding questions relevant to global political, social, and economic issues. A similar project called \href{https://en.wikipedia.org/wiki/Metaculus}{Metaculus} \cite{metaculus} includes a very broad range of topics. It also uses a sophisticated set of tools for analyzing aggregated predictions. Both GJO and Metaculus use \href{https://en.wikipedia.org/wiki/Scoring_rule#Proper_scoring_rules}{proper scoring} rules to record and aggregate numerical measures of forecast calibration. 
As will be seen, my proposal is very similar to GJO and Metaculus, but it will be oriented around much more technical and specialist-oriented questions. 
Prediction tournament methods have also been applied in more technical venues like the community-wide \href{https://predictioncenter.org}{CASP (Critical Assessment of Structure Prediction)} and \href{https://www.capri-docking.org}{CAPRI (Critical Assessment of PRedicted Interactions)} challenges for protein structures and protein-protein docking predictions, respectively. 

Traditional prediction tournaments like GJO, Metaculus and others are limited by the fact that some form of central authority or referee is necessary for deciding the outcome of predictions, and this requires a high level of trust on the part of participants. 
That is not normally a significant problem for the types of questions one often finds in these prediction competitions. They are relatively easy to state and the predictions (usually) have reasonably clear resolutions. For instance, a question such as “will country X invade country Y before the end of the 2022?” will almost always have a totally unambiguous answer. In general, therefore, the participants can rely on those running the competition to understand and fairly resolve prediction outcomes. 

However, with highly technical sciences the situation is rarely ever this clear cut, and scientists regularly fail to reach consensus even over the basic question of whether research actually supported or challenged a particular set of theories or assumption. Those with the technical expertise necessary to judge prediction outcomes are often part of small groups of specialists with their own biases and motivations. Therefore, even scientists who agree that improved methods for assessing predictions are badly needed will generally be dissuaded from participating in systematized prediction competitions, given that the accuracy of their predictions will be judged and assessed by their competitors. 

If prediction-tracking is to become a useful strategy for confronting the problems listed at the beginning of this section, then it is clear that there is a need for more generally applicable methods for assessing predictive claims in highly technical scientific subjects. Participants must be able to trust that
prediction assessments are legitimate and that they are judged by reasonably objective standards. Otherwise, they are unlikely to participate. 
Likewise, outside observers must be able to trust that claims of predictive power represent the true accumulation of scientific knowledge and are not just the artifacts of group-think or social reinforcement.
Of course, deferring to scientific expertise will always necessarily play an important role in assessing the outcome of any predictions.
So, the main difficulties lie with the paradox of designing a system meant to mitigate scientists' biases and misaligned incentives by consulting those very same scientists. That paradox is one of the main problems my proposal is meant to address.

In building a system that uses predictions to measure scientific progress, the challenge is therefore to seek out strategies that amplify the signal of expert scientific wisdom while filtering out the noise created by all the problems listed throughout this section. The problem with many existing systems for disseminating scientific information is that they frequently achieve the reverse.  
There are few incentives for scientists to state prior expectations clearly and in a way that can later be evaluated by outsiders. There are fewer still to openly acknowledge instances where expectations are overturned.

\section{A prediction-based consensus algorithm}
\label{s.predictions}

There are analogies between the challenges of prediction-assessment that I just discussed and the governance problems that \href{https://en.wikipedia.org/wiki/Distributed_ledger}{distributed electronic ledger} technologies like \href{https://en.wikipedia.org/wiki/Blockchain}{blockchains} are designed to deal with. In the latter case, groups of semi-anonymous actors need ways to participate in high-stakes financial transactions without a guarantee of absolute trustworthiness on the part of every participant, and where information available to some participants might be hidden to others. 
Part of the solution devised by cryptographers is to use clever consensus algorithms like \href{https://en.wikipedia.org/wiki/Proof_of_work}{proof-of-work} or \href{https://en.wikipedia.org/wiki/Proof_of_stake}{proof-of-stake} protocols. All such schemes involve some form of competition which, through checks and balances, preserves the information integrity of individual transactions. The labor that miners or stakeholders perform in attempting to extract financial rewards also functions to safeguard the integrity and the accuracy of information on a blockchain. In other words, individual incentives promote collective reliability and trust. 

My proposal takes an analogous balance-of-incentives strategy toward promoting information integrity in a database of scientific predictions. The system is to be organized around questions and predictions (and consensus about those predictions) rather than financial transactions. The goal is to harness the technical expertise, self-interest, and competition that exists within a group of scientists in a way that maintains the integrity of a database of prediction assessments. It is also to be designed in such a way that the processes of posing scientific questions, making predictions, and assessing their outcomes are transparent to any observer, even in disciplines that are impenetrable to most outsiders. 

In the next few sections, I will describe my own ideas for such a protocol. Its guiding principle is that valuable, but frequently lost, information is contained within the following:
\begin{itemize}
\item The extent to which members of a scientific community agree or disagree about a theory, model or hypothesis \emph{before} the relevant research projects to test it are performed,
\\

\item The exact extent to which pre-existing theories, models or hypotheses are actually involved in concrete predictions, 
\\

\item The extent to which members of a scientific community reach consensus over whether those predictions were confirmed or refuted \emph{by} the subsequent research and,
\\

\item The extent to which the community of scientists updates or modifies its confidence in theories, models and/or hypotheses based on prediction outcomes \emph{after} research projects are completed.
\end{itemize}
In much of current practice, these details are not logged or recorded in any quantifiable or systematic way while research is actually being conducted. To the extent that collective reasoning about predictions can be reconstructed later on, it is often only through post hoc and largely subjective studies of past published literature. 

My proposed consensus algorithm is meant to incentivize scientists to build a log of all this activity automatically. The goal is for the information stored in the log to be both transparent and accessible, not only to scientists' peers but also to outside observers. The experts themselves will perform the labor of checking that questions are well-posed and of ensuring that prediction outcomes can be (and are) evaluated in the most objective ways possible. As I will discuss in the coming sections, a system of checks and balances is used to limit the impact of any distortions brought on by the problems I discussed in section~\ref{s.problem}. In principle, any future observers should be able to search a log of questions and predictions to understand how confirmatory scientific research guided a scientific community toward its state of consensus and/or disagreement. 

The system I will describe in the next few sections works by distributing reputational award points based on the following activities on the part of participants:
\begin{itemize}
\item Formulating nontrivial and interesting questions about future events whose outcomes might be determined by subsequent scientific research efforts,
\\

\item Submitting predictions about the outcomes of those events,
\\

\item Reaching consensus about the outcomes of those predictions. That is, do scientists agree with their peers that a research project confirmed or refuted a particular prediction?
\end{itemize}
In its final form, the algorithm is to be integrated into a searchable web-based database, which I am calling \textit{Ex Quaerum} (``from the questions''). 

\section{\textit{Ex Quaerum}}
\label{ex_quaerum}

So far, I have discussed only abstract goals and the
general principles that motivate them. 
In the last section I described a kind of idealized question-prediction-resolution database.   
This paper will culminate in section~\ref{s.algorithm} with a description of an actual quantitative realization of such an algorithm. It will take the form of an award distribution algorithm aligned with the principles in section~\ref{s.predictions}. 

But since the system involves multiple intersecting components, and the incentive structure it is designed to create is somewhat complex, I will build up to it very slowly. 
I will begin in this section by describing in purely qualitative terms how I envision the steps of posing questions, submitting predictions, and reaching consensus, and I will start by using highly over-simplified scenarios as examples. The purpose here is to provide the reader with a very basic initial intuition for how the system I have in mind is meant to work, leaving aside the details and the mathematical formulas for section~\ref{s.algorithm}.

Within the \textit{Ex Quaerum} database, it will be useful to group users into two types: 
\begin{enumerate}

\item \underline{Expert Users}: These users will register to the system with unique user IDs. They consist of the expert scientists who actually pose questions, submit predictions, and work with their peers to validate prediction outcomes. Based on their interactions with the system, they try to accumulate reputational reward points. 

\item	\underline{Public Users}: Any member of the general public can use \textit{Ex Quaerum} to search for statistical information regarding questions, predictions and their outcomes. Searches can be performed for specific projects (a particle collider experiment, for example), user IDs, or even general scientific fields through the use of keywords. Searches will return metrics summarizing both predictive accuracy and expert consensus.  

\end{enumerate}
Since it is the expert users who have the more complex interactions with the system, they will dominate the discussion for the remainder of this paper. 

\subsection{A single question, prediction, and resolution sequence}
\label{single}


Each individual expert registers to the system with a unique ID. (Hereafter, I will refer to any expert user who interacts with the system simply as a “user.”)  The users submit questions about future developments with the goal of attracting other experts to submit predictions about the outcomes. To keep the discussion simple, I will assume that all questions come with yes-or-no answers. Expert users have three basic tasks: 

\begin{enumerate}
\item To propose the questions about future events whose outcomes are to be predicted. These could be questions about, for example, the outcomes of specific experiments. \vspace{.1in}

\item To make and submit the predictions. \vspace{.1in}

\item To confirm prediction outcomes and work with other expert users to reach consensus about those outcomes.
\end{enumerate}

As an (entirely hypothetical) example, a climate scientist might submit the following question for consideration:
\bquote{Recently there has been debate over the ability of current climate models to anticipate rates of large storms in the US. Reference A seems to indicate that the evolution in the rate of large storms will deviate wildly from the predictions of reference B which are based on more standard modeling techniques. Will the model of reference A provide a more accurate estimate than reference B of the number of large storms observed in the US in the 2020s?}
Let us call this “Question A.” It is logged into the \textit{Ex Quaerum} system and given a tentative label with the date and subject category (in this case, ‘tornado science’). 
\begin{figure*}
\centering
\includegraphics[scale=0.75]{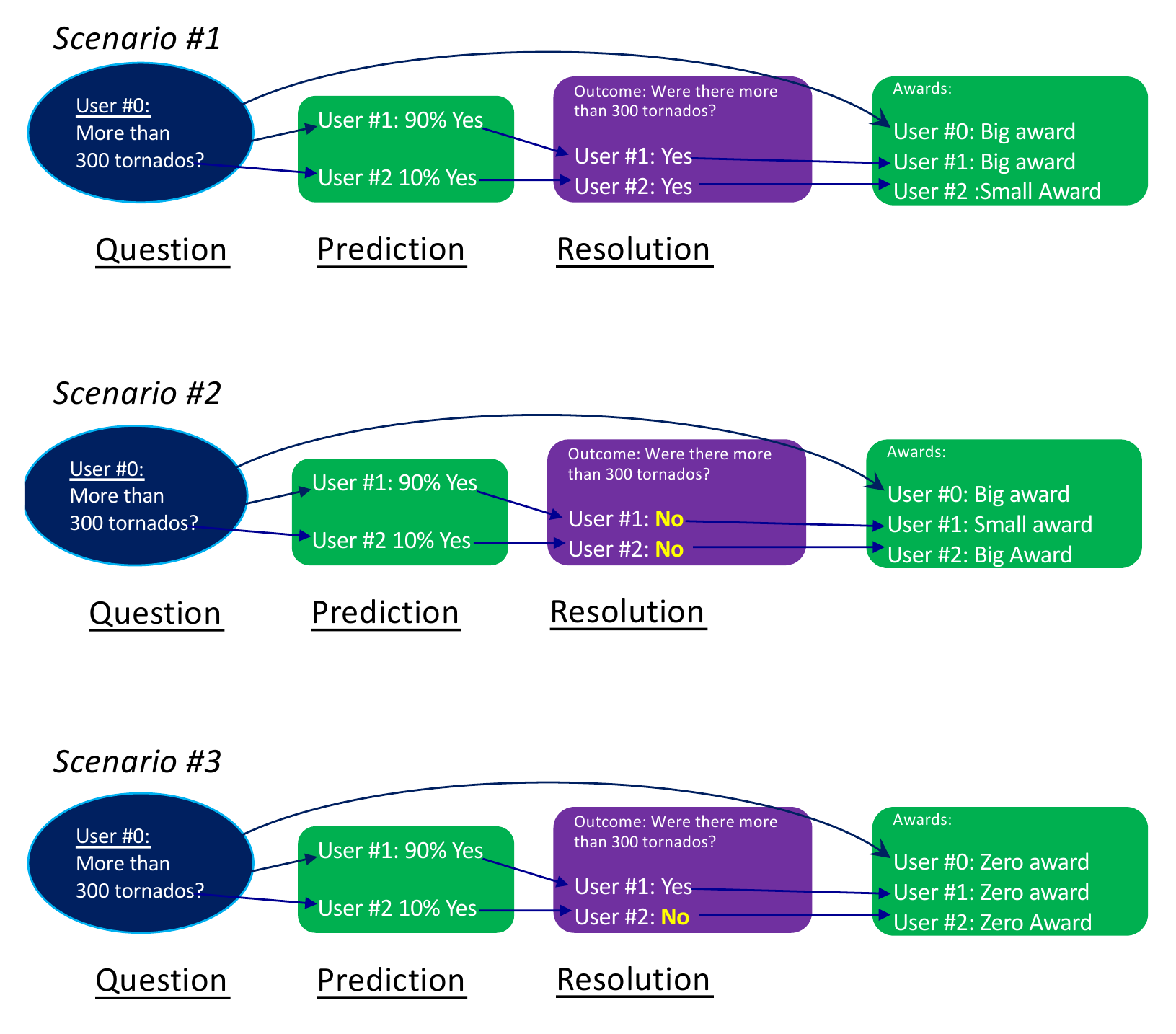}
\caption{The elements of a question-prediction-resolution sequence for a single question and with two expert users submitting predictions. The three possibilities discussed in the text are shown as scenarios \#1 through \#3.}
\label{f.scenarios1th3}
\end{figure*}
\begin{figure*}
\centering
\includegraphics[scale=0.75]{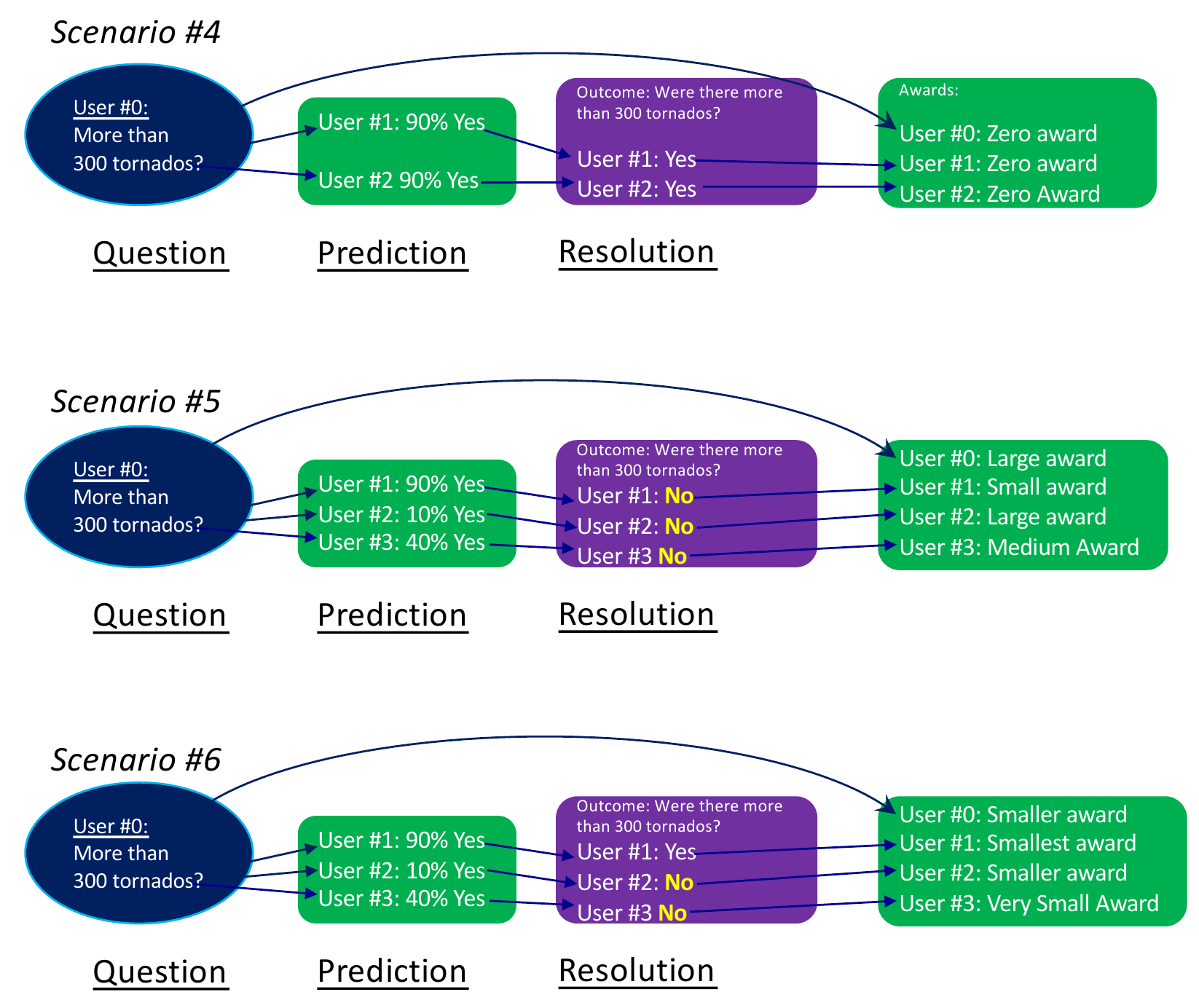}
\caption{Like figure \ref{f.scenarios1th3} but now for scenarios \#4 through \#6 in the text.}
\label{f.scenarios4th6}
\end{figure*}

In the next stage of the process, other users respond by submitting predictions in the form of a probabilistic forecast for a “yes” result. To see how this works in the above example, imagine that there is a group of researchers associated with reference A located at a university in Stockholm and another group of researchers associated with reference B in Chicago. Each group would like to submit a prediction that proves that their model, and with it a set of underlying assumptions, is the superior one. 

However, as it is stated, Question A is far too vague to provide much unambiguous guidance. Therefore, the users will likely enter into negotiations to clarify the question to a point where both groups are willing to submit predictions. After some back-and-forth discussion,
the user who originally posed the question might update it to something like:
\bquote{Recently there has been debate over the ability of current climate models to anticipate rates of large storms in the US. Reference A seems to indicate that the evolution in the rate of large storms will deviate wildly from the predictions of reference B which are based on more standard modeling techniques. For example, the simulation in reference A seems to allow for as many as 300 tornadoes in Kansas in 2024, while such a large number is excluded in the simulation used by reference B.  Will the number of tornadoes observed in Kansas in 2024 equal or exceed 300?}
It might be necessary for the users to refine the question phrasing even further before enough people are willing to submit predictions.\footnote{I am making the example with tornadoes intentionally silly. I do not have a background in either climate science or meteorology, and I know nothing about the science of tornadoes. Nor do I know anything about any climate science work being done in Chicago or Stockholm. My purpose here is to use a vivid narrative for illustration purposes while avoiding distractions from actual scientific questions and controversies.} That might mean clarifying the exact technical details of how measurements are to be performed, what published sources are acceptable for resolving the outcome, methods for dealing with statistical uncertainties, and so forth. Ultimately, however, users from the Chicago and Stockholm groups agree to a version of Question A and they are ready to submit predictions to be recorded to the system.

To simplify this illustration still further, assume for now that there are only two users making predictions who I will call user \#1 and user \#2. User \#1 is from the Stockholm group and user \#2 is from the Chicago group. 

User \#1 is convinced that the climate model from the Chicago group uses badly out-of-date assumptions. Based on their “improved” Stockholm model, user \#1 believes it is almost a certainty that there will be more than 300 tornadoes in Kansas in 2024. User \#1 therefore submits a probability of 90\% that the answer to Question A is “yes.” 

Conversely, user \#2 thinks the Stockholm group has misunderstood important fundamentals of climate science. They argue that their Chicago model is based on years of careful development and that, unlike the Stockholm model, it has been verified many times by past data. They emphasize that it has been tuned to careful measurements over a period of decades. The Chicago model gives a likelihood of 300 or more tornadoes that is very small. Therefore, user \#2 submits a forecast of only 10\% that the answer to question A is “yes.” In addition to submitting their probabilistic predictions, each user is also permitted (and encouraged) to submit a short explanation of their reasoning. Both the forecasts and their accompanying explanations are dated. 

Now these predictions (and explanations) are stored in the \textit{Ex Quaerum} system and left there until 2024. After 2024, users \#1 and \#2 return and indicate whether they agree that there actually were or were not more than 300 tornadoes in Kansas in 2024. If both agree that the answer was “yes,” then user \#1 is recorded as having the more highly calibrated prediction. User \#1 thus receives a positive number of reward points, while user \#2 receives little or none. This sequence of events is summarized graphically under \emph{scenario \#1} of figure~\ref{f.scenarios1th3}. If both user \#1 and user \#2 answer “no,” then it is user \#2 who receives the positive points while user \#1 who receives the small award. This is shown as~\emph{scenario \#2} of figure~\ref{f.scenarios1th3}.

If one user declares the outcome to be “yes” (there were 300 tornadoes or more in Kansas in 2024) but the other insists it was “no,” then there is no prediction consensus. In that case, no award points are distributed to either user and the question is recorded in \textit{Ex Quaerum} as a failure. This is illustrated under \emph{scenario \#3} of figure~\ref{f.scenarios1th3}.

I will call the user who submitted the original question “user \#0.” (Note that this means that there are actually three users involved: one who proposed the question and two who submitted predictions.) In figure~\ref{f.scenarios1th3}, user \#0 is represented by the dark blue ovals on the left. If a consensus is reached on question A, then user \#0 receives a large number of reward points as payment for guiding the question through to a successful resolution. If Question A fails, user \#0 gets no reward points at all. 

Now consider what a member of the general public sees if they search for information about Question A and find either of the first two scenarios. They will likely notice that two experts with initially opposing views nevertheless reached the \emph{same} conclusion about the outcome of the prediction. They can reasonably infer, therefore, that the evidence for the prediction outcome was at least compelling enough to convince one of the users of its truth despite having to accept a losing position and a small award.
That is, any quoted research must have been convincing enough to change minds.
Moreover, the question must have been interesting, relevant, and topical enough to draw the attention of at least two competitive experts.  

If instead the member of the general public searches on \textit{Ex Quaerum} and finds that scenario \#3 was the outcome, then they can see that the research apparently had little influence on the two participating experts. To outside observers especially, this prediction competition has clarified little or nothing because it remains unclear whose expectation was closer to an objective outcome. For a ``failed'' question like this, all users get zero reward points.  

Since they are required to convince competitors about outcomes in order avoid failed questions, users in this system are incentivized to state assumptions clearly and to make agreements in advance about what exactly constitutes a ``yes'' or ``no'' outcome for each question. The need for \emph{both} competitors to be convinced also disincentivizes activities like
goalpost-adjusting. 

 For the questioner, user \#0, there is also a reward incentive to pose the question clearly enough that there can be no ambiguity about the prediction outcome. This user must also make the question interesting enough to attract predictors. If the question is so sloppily posed that users \#1 and \#2 are unlikely to ever reach consensus about outcomes, then user \#0 is also unlikely to ever collect a reward. 
 
 In a broader sense, beyond the reward distribution, all three of the participating experts in the above example stand to benefit collectively from the final success of the question if it demonstrates good predictive power in their particular field.

Also notice that all the participants are incentivized to carefully scrutinize any relevant research. If it is purported that some new research result determines a particular prediction outcome, then at least one of the users will need to accept a losing position in order for the question to be considered successful. That user will of course be incentivized to study the research results especially carefully and call out any missteps; of course, they will not be inclined to concede to having lost a prediction competition unless they are compelled to by truly convincing research. Having more participants in \textit{Ex Quaerum} means there are more researchers doing the work of checking and validating, possibly even to the point of replicating or reproducing, scientific research.  

It is worth emphasizing this point further: Consider the role of any other scientists, involved with \textit{Ex Quaerum} or not, who might be performing detailed, maybe highly technical research that impacts the resolution to the debate over the Stockholm versus Chicago perspectives. A concern along the lines of what was discussed in the introduction is that the research might be biased in one direction or another due to the researchers' prior beliefs, but that the problems might go unnoticed due to the technical nature of the work. But it is not possible for any relevant research to completely avoid skeptical scrutiny by expert users in \emph{Ex Quaerum}. In all the scenarios, the users have opposing views. So, in whichever biased direction the scientists conducting research might lean, either user \#1 or user \#2 will have a stake in the opposite perspective. Thus, the interests of competing scientists are  balanced against one another in a way that promotes careful scrutiny of the research.

To summarize, the incentives intersect to promote well-posed questions, unambiguous conditions for their resolution, skeptical scrutiny of relevant research, and consensus about the final prediction outcomes. 

But consider a fourth scenario where both user \#1 and user \#2 submit exactly the same prediction. Say, for example, that when user \#2 runs the Chicago simulation they actually find a fairly high chance of at least 300 tornadoes appearing in Kansas in 2024, so they also submit a 90\% “yes” forecast along with user \#1's 90\% forecast. This situation would completely undermine the incentive structure we described above in the previous three scenarios. 
Thus, if all predictions on a question are identical, the question will again be recorded as a \emph{failure} in \textit{Ex Quaerum} and no reward points will be distributed. The diagram for this is shown as scenario \#4 in figure~\ref{f.scenarios4th6}. Again, the reason for a zero reward payout in a question with identical predictions is that the system of checks and balances that existed in scenarios \#1-\#3 by the competing viewpoints has been lost. 
No user is especially incentivized to question the relevant research. 

The zero reward payout for questions like scenario \#4 eliminates some of the strategies participants might be tempted to use to exploit or game the system. For example, cognitive biases might lead a group of scientists to overestimate (probably subconsciously) its predictive power and thereby confirm a prediction despite indications that it might have failed. It is also simple to imagine groups of users submitting predictions on easy but irrelevant questions as a way to inflate their apparent predictive abilities. But in the above system, it is not possible to gain reward points in this way. The users who participate on a given question \emph{must} provide opposing predictions for a question to ultimately be recorded as a success and for any reward points to be distributed to any of the participants.

One can also imagine a strategy to game the system that is the mirror image of the one just described in the previous paragraph, wherein a group of scientists collude (again, probably subconsciously) to submit equal but intentionally \emph{poor} forecasts. The motivation in this case would not be to showcase predictive power but rather to highlight how surprising, and therefore interesting or useful, a particular research project might be. Again, the zero reward for equal forecasts eliminates this as an effective gaming strategy in \emph{Ex Quaerum}. 

To summarize, the wider the difference there is between different users' predictions, the stronger is the motivation for at least one of them to carefully examine the details of scientific research before agreeing to validate a particular prediction outcome. To an outside observer, an outcome like scenario \#4 reveals that no concrete scientific debates were resolved by any of the relevant research, and the reward algorithm in \emph{Ex Quaerum} is designed to reflect this. 

In order to receive a potentially large number of reward points, user \#0 must work to avoid outcomes like scenario \#4, so they are incentivized to seek out questions that are both interesting and relevant enough to generate genuine debates and attract a range of competing yet concrete opinions. 

To further understand how the system motivates transparency, 
it is helpful to imagine the type of situation that might lead to a scenario \#3 in the tornado narrative. Say that the method for detecting tornadoes involves distributing expensive tornado detecting towers filled with electronic equipment. The group performing the measurements in 2024 has a limited budget and needs to be judicious in how it decides to place these towers across the open spaces of Kansas. After a careful statistical analysis, they conclude that the most useful data will be obtained if detection towers are mostly placed in areas that are already expected to have many tornadoes. However, to estimate the relative frequency of tornadoes in different parts of the state, they have to use simulations. For various technical reasons, they decided to use simulations from the Chicago model. 

Now, there might be a legitimate concern that the Chicago model is influencing the very data that are ultimately going to be used to test it. But the researchers might also have found very compelling arguments that have convinced them that using simulations in this way will not significantly influence their final results. 

Regardless, in this story the data are taken, analyzed, and published following the above methodology, and they seem to show that there were fewer than 300 tornadoes. User \#2 thus submits a ``no'' at the resolution stage and claims predictive victory. However, the Stockholm group immediately objects, and highlights the fact that the Chicago model actually guided the data-taking process that the researchers used. They make a convincing  case that this is methodologically flawed, that the analysis was excessively biased, and they also claim to have made their own estimates suggesting that if the researchers had placed detection towers in accordance with the Stockholm model then they would, in fact, have measured more than 300 tornadoes. 

The researchers who did the tornado measurements respond to this criticism. They claim that, prior to beginning their work, they actually \emph{did} compare the ways that the Chicago and Stockholm models would affect detection tower placement, and that they found that different tower placements would have resulted in a negligible effect. They explain that they published their plots and associated discussion of the Chicago model's estimates alone only because it was easier to run the simulation quickly at a practical level. The Stockholm group is still unconvinced by this explanation. They respond again and disagree with a number of technical points in the way their simulation was compared with the Chicago simulation. They insist that if their model had been used correctly it would have led to a dramatically different tornado tower placement and an outcome with more than 300 tornadoes.  They therefore insist on a ``yes'' at the resolution stage. The debate gets highly technical from this point, and it starts to involve esoteric statistics, computer codes, etc. It also gets heated, with the Chicago group accusing the Stockholm researchers of refusing to acknowledge problems with their model in the face of concrete data, and the Stockholm group accusing the Chicago group of baking confirmation bias into their data-taking procedure. 

Once a debate like this spirals beyond a certain point, it becomes unclear to anyone looking on from the outside what the true outcome was, at least without devoting  unrealistic amounts of time and effort to unpacking all the details. 
To anyone who was not already a Chicago or Stockholm adherent from the outset, the situation is just confusing.
Overall, therefore, the research did not successfully contribute very much to increasing general, widespread and agreed-upon understanding of tornadoes. Thus, this scenario \#3 results in zero award being distributed to any of the scientists involved. Ideally, all of the scientists involved in the tornado research should have debated the methodological issues and reached a reasonable level of agreement about the appropriate use of simulations \emph{prior} to the research being performed. 

It is worth emphasizing that debates over the legitimate predictive success of scientific ideas are very common in practice, especially in very highly technical fields. For a very public example, see the critique of inflationary cosmology here~\cite{ijjas2017pop} and the response here~\cite{inflaction_response}. This is an example where careful records of precise predictions, the reasoning behind them, and their outcomes would have been helpful, especially to onlookers.

Scenarios \#3 and \#4 both result in zero reward for participants, but for opposite reasons. For any participants to get a high reward, there must be a diversity of predictions before the research, but consensus about outcomes afterwards. 

\subsection{More users and questions}
\label{three}

Very little can be learned with any confidence from a single bet between only two individuals. Furthermore, there will almost never be absolute, universal agreement about outcomes. The purpose of the above examples is only to explain the basic strategy of the system. To be statistically meaningful, there should be many users submitting many questions, predictions and resolutions. 

Therefore, to further build intuition for how the system will work in an actual implementation, consider how the basic examples in section \ref{single} change if there is a third user (user \#3) also submitting predictions. For example, say there is a researcher in Tokyo watching the dispute between users \#1 and \#2. User \#3 basically agrees with user \#2 that user \#1 is ignoring too much of the standard work on climate science, but user \#3 also thinks user \#2 is overly dismissive of new ideas from the Stockholm group that do show some promise. In user \#3’s estimation, those new ideas have at least some chance of eventually replacing the old ones once they have been properly refined. User \#3 therefore submits a 40\% “yes” prediction along with an explanation of their reasoning. 

If fewer than 300 tornadoes are observed, and all users agree on this outcome, then the resulting chart will be what is shown under scenario \#5 in figure \ref{f.scenarios4th6}. 
Since user \#3 is closer to user \#2, and user \#2’s forecast is more closely calibrated with the consensus result, then user \#3 receives a reasonably large number or reward points, though not quite as many as user \#2.

But now imagine that user \#1 does not agree that users \#2 and \#3 are interpreting the new research outcomes correctly. User \#1 believes that there actually \emph{were} more than 300 tornadoes in 2024. They might argue, for example, that instruments for recording the presence of a tornado were not properly calibrated or that the definition of a tornado used by users \#2 and \#3 is too narrow. The participants might be in a state of conflict similar to what was described at the end of the last section. User \#1 thus insists on a “yes” result in the resolution stage of the question, breaking with users \#2 and \#3. The diagram for this is scenario \#6 in figure 
\ref{f.scenarios4th6}.
There is not a total loss of consensus in scenario \#6 because 2/3 of the users still agree on the outcome, but now the consensus is weaker. User \#3 has essentially broken the tie in debate described at the end of the last section. This question outcome is still valuable, but perhaps less so than if all expert users were in unanimous agreement about the outcome, and this state of partial consensus should be reflected in the reward distribution as the diagram shows. User \#2 has been given the largest reward since their forecast is most closely aligned with the average consensus, but their reward is smaller than it would have been in scenario \#5 where the consensus was unanimous. 
When an external observer finds question A in \textit{Ex Quaerum} in a search, they get a sense of both the average expert confidence \emph{and} of their tendency to reach consensus after the fact.  

\begin{figure*}
\centering
\includegraphics[scale=0.5]{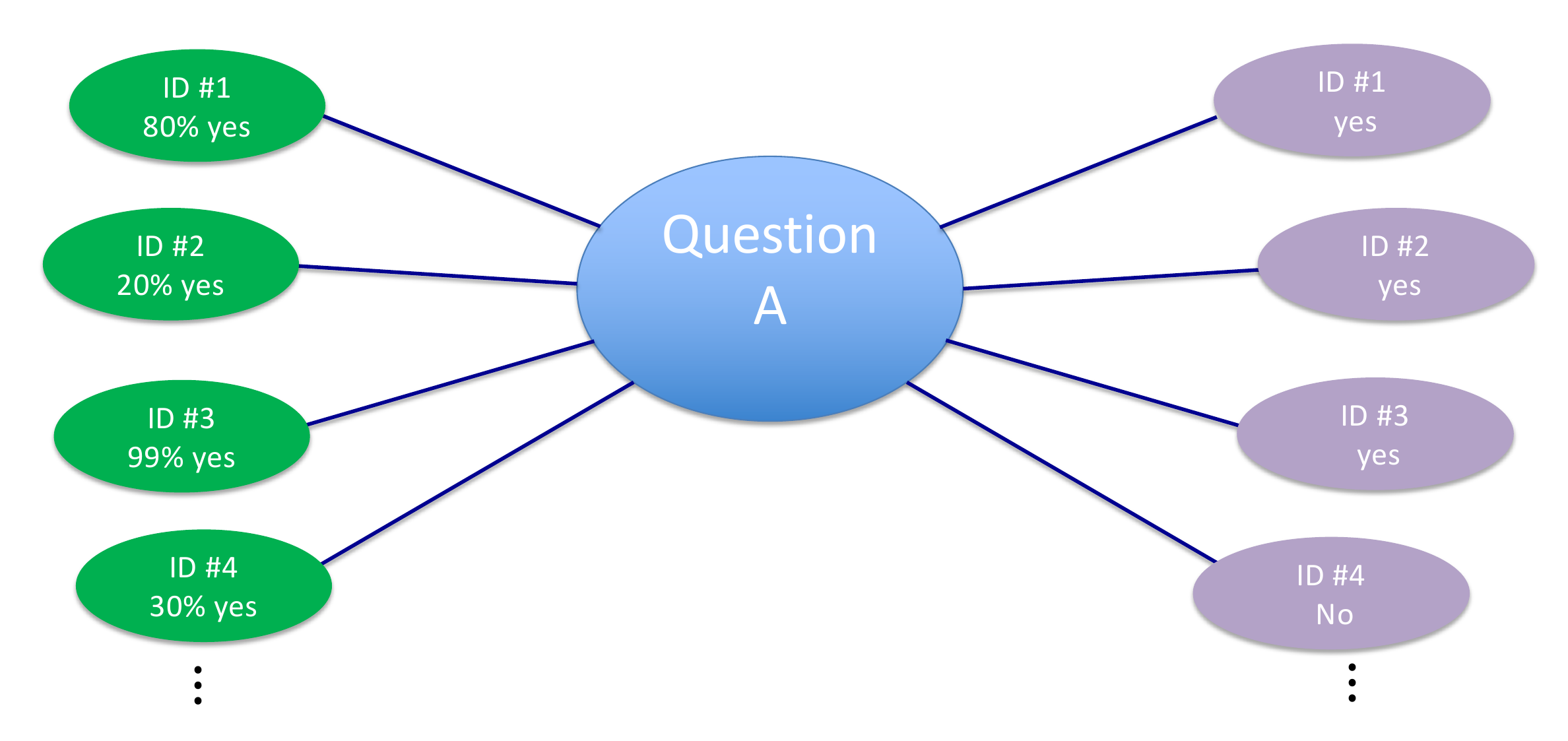}
\caption{A diagrammatic representation of a single question with many expert user predictions. The central blue oval labels the question. The green ovals on the left represent the yes/no predictions with the ``ID \#'' indicating each user. The violet bubbles on the right are the corresponding claimed outcomes from each expert user. (See text for further explanation.)}
\label{f.question}
\end{figure*}

Of course, examples, with only three predicting users on one question, are still far too limited to be statistically interesting. 
For the system to be meaningful there need to be many questions and many users per question. But it should still simultaneously reward predictive accuracy, a wide range of predictions, and ultimate consensus in a manner similar to what I illustrated above. In a genuinely interesting scenario for a single question, a visualization diagram  might look something like figure~\ref{f.question}.
The central blue oval represents an original question posed by one of the expert users (corresponding to user \#0 in our two and three user examples above).
The green ovals on the left of that diagram represent the predictions from different expert users and the violet ovals on the right represent their corresponding resolution statements. The vertical “…”s indicate that there are many additional predicting users. For each oval there may be textual content discussing the reasoning behind the question/prediction/resolution and including links to external literature.

\begin{figure*}
\centering
\includegraphics[scale=0.5]{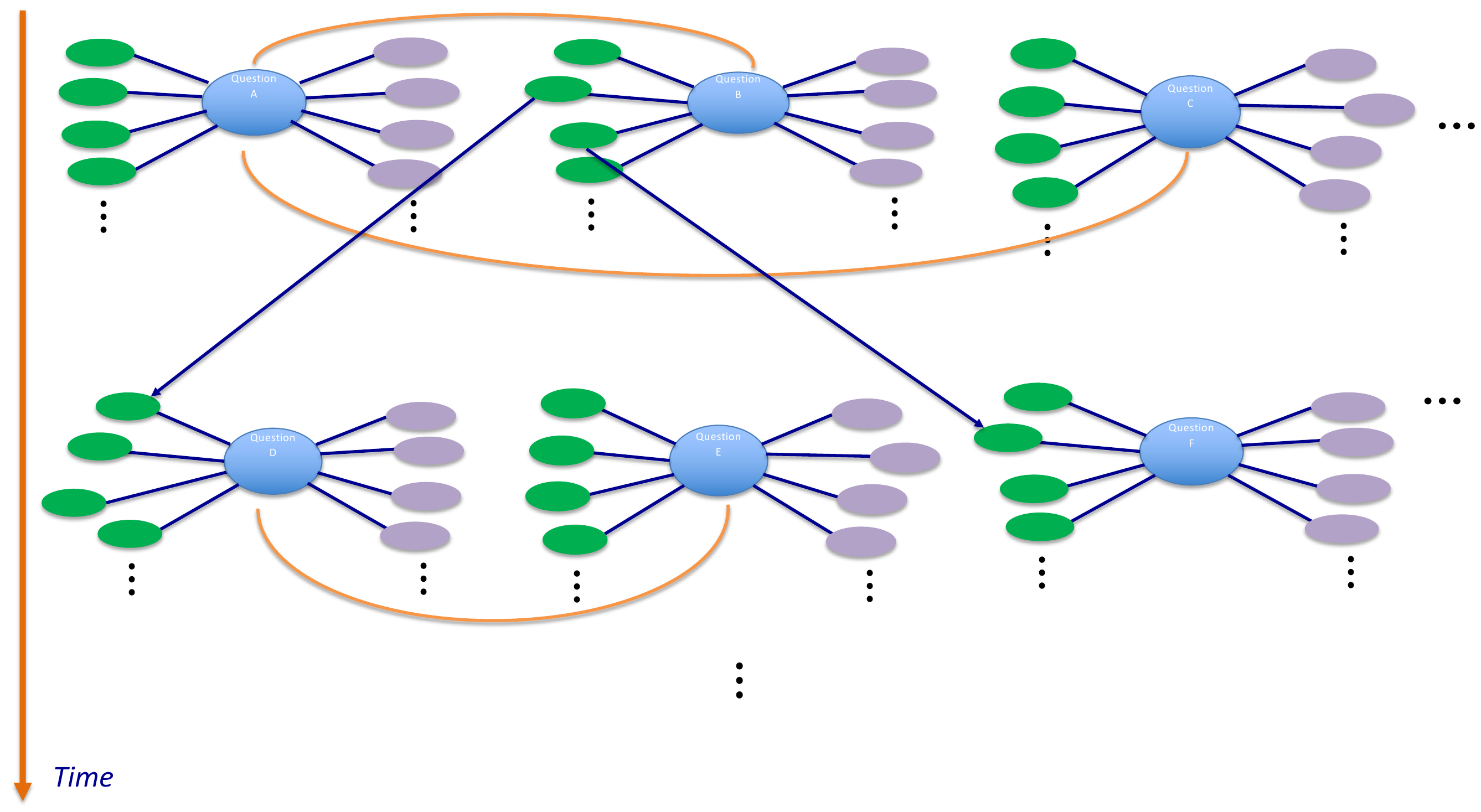}
\caption{Over time, questions like figure~\ref{f.question} populate the \textit{Ex Quaerum} database. An external observer can search for connections between the topics and literature used in different questions and by different users, indicated here by the blue and orange lines.}
\label{f.many}
\end{figure*}

Within \textit{Ex Quaerum}, these questions are to be organized into a database with connections between different questions indicating relationships by topic and quoted literature. The full database might be visualized as in figure~\ref{f.many}.
Questions and predictions are linked within the database according to common themes as indicated in 
figure~\ref{f.many} by blue and orange lines. 
Anyone can search for metadata pertaining to specific features like subject, associated theoretical assumptions, literature cited, or other relevant information provided in the database. 

Up to this point, my purpose has been to explain the general strategy and philosophy. It should be clear by now that the basic technical challenge 
is to determine exactly how to distribute award points based on the logic of section~\ref{s.predictions} and illustrated in the (highly oversimplified) scenarios in section~\ref{ex_quaerum}. In several places I referred to users receiving "large," "small," or "medium" amounts of reward points. Making this quantitative requires a specific algorithm for scoring forecasts and distributing award points. That is what I will turn to next.

\section{Scoring and reward points}
\label{s.algorithm}

Methods for numerically scoring forecasts have a long history --  for a popular account, see for example   Ref.~\cite{tetlock_superforecasting_2016}. Their purpose is normally to reward most highly those probabilistic forecasts that most closely match the actual frequency of events.\footnote{For example, over a long period of time it should rain on about 20\% of the days that a weather forecaster forecasts a 20\% chance of rain.} But the goal of \textit{Ex Quaerum} is not only to incentivize forecast calibration alone, but rather to motivate the whole spectrum of good practices outlined in section~\ref{s.predictions}, including the formulating of clearly stated questions that can resolve open debates, the reaching of consensus about research outcomes, and the inclusion of a wide range of viewpoints. To combine all those elements, something more than a straightforward forecast scoring rule is necessary. 

Therefore, I will treat scoring prediction accuracy and distributing reward points as two distinct but related tasks. I will explain my proposed mathematical formulas for each in the next two subsections, starting with the prediction scoring algorithm.

Before beginning, some definitions are necessary: 
\begin{itemize}

\item 
$r_{i,j}=$ the number of reward points given to user $i$ on question $j$.
\vspace{.1in}

\item 
$N_j=$ the \underline{number of users} who submit predictions for question $j$. In the section~\ref{single} examples involving tornadoes (figure~\ref{f.scenarios1th3}), $N_{j=A}=2$. 
\vspace{.1in}

\item $p_{i,j} = $ the \underline{probabilistic prediction} (forecast) supplied by user $i$ on question $i$. For the first tornado examples of section~\ref{single} and figure~\ref{f.scenarios1th3}, $p_{1,A}=0.9$ and $p_{2,A}=0.1$. \vspace{.1in}

\item $v_{i,j} = $ the \underline{resolution} of question $j$ asserted by user $i$: \\ 
\begin{tabular}{lll}
& $v_{i,j} = 1$ & if user $i$ says the outcome is ``yes'' \\
& $v_{i,j} = -1$& if user $i$ says the outcome is  ``no'' \\
& $v_{i,j} = 0$ & if user $i$ supplies no answer \, .
\end{tabular}
These $v$'s correspond to the third bubbles from the left in figures~\ref{f.scenarios1th3} and~\ref{f.scenarios4th6}.
\vspace{.1in}

\item $V_j = $ the \underline{mean resolution} of all users on question $j$.
\vspace{.1in}

\item $q_j = $ the \underline{outcome} of question $j$:\\
\begin{tabular}{lll}
& $q_j = 1$ & for ``yes,'' \\
& $q_j = -1$ &  for ``no,'' \\ 
& $q_j = 0$  & for ``unresolved.'' 
\end{tabular}
\end{itemize}
The last two quantities are to be calculated from the previous three as follows: First, $V_j$ is just the mean of  
all the asserted outcomes to question $j$, averaged over all users,
\begin{equation}
\label{e.meanV}
V_j = \frac{1}{N_j} \sum_i^{N_j} v_{i,j} \, .
\end{equation}
The question outcome $q_j$ is ``yes'' for question $j$ if the majority of users 
declare it was ``yes'' and ``no'' if the majority of users 
declare it was ``no.'' That is, 
\begin{equation}
q_j = \begin{cases}
	+1 \,& \text{if} \;\;  V_j > 0 \\
	-1 \, & \text{if} \;\;  V_j < 0 \\
	0 \, & \text{if} \;\;  V_j = 0 \, 
\end{cases} \, .  \label{e.outcome_def}
\end{equation}
To see some examples, return to the scenarios of the previous section. Scenario \#1 in figure~\ref{f.scenarios1th3} has $V_A=1$ with $q_A=1$, scenario \#2 has $V_A=-1$ with $q_A=-1$, 
scenario \#3 has $V_A=0$ with $q_A=0$, scenario \#4 of figure~\ref{f.scenarios4th6} has $V_A=1$ with $q_A=1$, scenario \#5 has $V_A=-1$ with $q_A=-1$, and scenario \#6 has $V_A = -1/3$ with $q_A = -1$.

I will call the absolute value $|V_j|$ the ``conensus'' on question $j$. $|V_j| = 1$ means there was unanimous agreement about the outcome, $|V_j| = 0$ means there was no agreement at all. In all of the above examples $|V_j|$ was either 0 or 1, complete success or complete failure, except in scenario \#6 where the consensus was $|V_j| = 1/3$. 

\subsection{Forecast scoring}

Prediction accuracy will be quantified using “surprisal,” which is a
\href{https://en.wikipedia.org/wiki/Scoring_rule#Strictly_proper_scoring_rules}{strictly proper} forecast scoring rule, meaning that users receive on average the best score by submitting their truest (or most “calibrated”) probability estimate given their state of knowledge. The definition of the surprisal given to user $i$ for  their forecast on question $j$ is
\begin{equation}
\label{e.surprisaldef}
s_{i,j} = \begin{cases} 
	-\ln p_{i,j} \, & \text{if} \, q_j = +1 \\
	-\ln (1-p_{i,j}) \, & \text{if} \, q_j = -1 \\
	0 \, & \text{if} \, q_j = 0
	\end{cases} \, .
\end{equation}
The goal of any user is to get the smallest possible surprisal. For example, in scenario \#1 of figure~\ref{f.scenarios1th3}, user \#1 said 90\% ``yes'' for a question that turned out to be “yes,” so their surprisal would be $\ln(0.9) \approx 0.11$. User \#2 said 10\%, so their surprisal would be $\ln(0.1) \approx 2.3$. Had the answer been “no,” as in scenario \#2, then the surprisals would be the same but reversed for the two users. A user who has no idea about the answer to a question might give a probability of 50\%, and their surprisal would then automatically be $\ln(0.5) \approx 0.69$ regardless of the outcome.  

Note that the surprisal, as we have defined it in equation~\ref{e.surprisaldef}, is not strictly for a forecast about what the actual or objectively ``true'' outcome of question $j$ will be, but rather for what the majority view, as expressed through equation~\ref{e.outcome_def}, will be. For now it should be considered to be an open question whether $q_j$ is a good proxy for the objective truth of the prediction outcome. 

There are other strictly proper scoring rules such as the Brier score, but surprisal has advantages that I hope will be clear in the examples I provide below.  
In a  sense, the magnitude of surprisal quantifies how “surprised” a user should be by the outcome of a prediction; a very confident prediction results in a very high surprisal in the case of the less expected outcome.~\footnote{See also \href{https://en.wikipedia.org/wiki/Information_content}{Shannon information content}.} 
A forecast of either 100\% or 0\% can result in infinite surprisal if the prediction fails -- discomfirmation of a belief about which one is infinitely certain results in infinite surprise.
Therefore, in calculations it will be necessary to put an upper limit on the allowable confidence of predictions. (Say, 99.9999\%.)

The range of surprisals between different users can be quite large even when all users have reasonably high confidence in the outcome. For example, say that the probabilities given for “yes” by three users on a particular question are 95\%, 99\%, and 99.9\%, but the outcome turns out to be “no.” The range of surprisals in this case are approximately 3, 4.6, and 11.5 respectively. Even though all three users basically agreed that the answer was very probably “yes,” the differences between their surprisal scores is significant. It is larger even than the difference between surprisal scores in a 90\% versus 10\% prediction contest, as in the scenarios of figure~\ref{f.scenarios1th3}. 

Once there is a database of forecast scores that involves many users and many questions, useful averages can be constructed. The mean $\langle s_j \rangle$ of scores from all users on question $j$ is
\begin{equation}
\label{e.meansur}
\langle s_j \rangle = \frac{1}{N_j} \sum_i^{N_j} s_{i,j} \, .
\end{equation}
Each user will strive to keep their average surprisal as low as possible.
In preparation for the next section, I also write the mean-squared surprisal,
\begin{equation}
\label{e.stdsur}
\langle s_j^2 \rangle = \frac{1}{N_j} \sum_i^{N_j} s_{i,j}^2 \, .
\end{equation}
The standard deviation in the surprisals on question $j$, calculated from equations \ref{e.meansur} and \ref{e.stdsur}, is a measure of their spread, or the amount of variation between different users' scores on that question,
\begin{equation}
\Delta s_j = \sqrt{\langle s_j^2 \rangle-\langle s_j \rangle^2} \, .
\end{equation}

On a given question $j$, it will ultimately be necessary to judge all users' level of surprise relative to one another. This will be needed for the reward distribution algorithm in the next subsection. A user $i$ is very surprised if their $s_{i,j}$ comes out significantly larger than what is typical for other users involved on question $j$. Within a group of users casting predictions for question $j$, I will consider surprisals more than a few standard deviations higher than the mean are to be unusually high. Thus, I define a ``big surprise'' on question $j$ to be 
\begin{equation}
\label{s.big}
s^{\text{Big}}_j = \langle s_j \rangle + c \Delta s_j \, ,
\end{equation}
where $c$ is a positive constant that can be tuned later on. It is simply the number of standard deviations above the mean surprisal that a user can have before it is considered a large departure from the group. A user with $s_{i,j} > s^{\text{Big}}_j$ is very surprised (relative to the other users) by the outcome. I will need to refer to $s^{\text{Big}}_j$ in the next subsection.   

\subsection{A reward distribution algorithm}

The reward point distribution algorithm is to formulated to 
achieve the incentive goals of section~\ref{s.predictions}. It will be rooted in the concept of the suprisal score described above, but there needs to be a reweighting to incentive i.) a large spread in surpisals and ii.) a final consensus close to $|V| \approx 1$. 
I will organize the explanation of the reward algorithm in two steps below. 

In the first step, I consider the case of a single user $i$ who always receives the same surprisal relative to the group,
\begin{equation}
\label{e.consts}
s^{\text{Big}}_j - s_{i,j} = \text{constant for all}\, j  \, ,
\end{equation}
on all questions. That is, their forecast accuracy is always the same relative to their peers for this set of question. 
Based on the discussion in section~\ref{s.predictions}, we must ask how many reward points this user should receive for each question. But since the user always has the same surprisal relative to the group, the accuracy of their forecasts does not provide an answer. Instead, their reward for each question $j$ should be based on whether that question had a large spread $\Delta s_j$ (spread) in suprisals and a large consensus $|V_j|$. 
Therefore, I propose to give this user a reward proportional to $|V_j|$ and to $\Delta s_j$. If $|V_j| = 1$ on question $j$ they receive a large reward, but if $|V_j|=.5$ they receive only half that reward because of the weaker consensus among experts about the outcome of question $j$. If $|V_j|=0$ there is of course no reward to distribute. Also, a nearly zero $\Delta s_j$ should result in almost no reward for this user since the checks and balances ensured by a spread in competing predictions are absent. The reward given out for each question grows larger as the spread $\Delta s_j$ grows. 
Since their reward is proportional to both $|V_j|$ and $\Delta s_j$, I define a special quantity 
\begin{equation}
\label{e.reward_scale}
R_j = \left| \Delta s_j V_j \right| \, ,
\end{equation}
which I will call this the ``reward scale'' because it sets a baseline relative to which rewards are distributed to all users.  

To summarize this first step, a user who always receives the same relative forecast score, equation \ref{e.consts}, will always receive a reward determined by, and proportional to, $R_j$. A large $R_j$ means there is both a wide spread of differing predictions and a broad consensus about the question's outcome. 

As a concrete example, in both scenario \#1 and scenario \#2 of figure \ref{f.scenarios1th3} there were scores of $-\ln{\left(0.9\right)} \approx 0.11$ and $-\ln{\left(0.1\right)} \approx 2.3$. So, $\Delta s_j \approx 1.1$ in both cases. Also in both scenario \#1 and scenario \#2, $\left| V_j \right|=1$. So, in both scenarios the reward scale is $R_j=\left| \Delta s_j V_j \right| \approx 1.1$. In scenario \#3, $R_j = 0$ because $V_j = 0$. In scenario \#4, $R_j = 0$ because $\Delta s_j = 0$. 

In the second step of the algorithm setup, I instead consider a large set of questions $j$ that all happen to end up with exactly the same $\Delta s_j$ and $|V_j|$ (and thus the same $R_j$ and $s^{\text{Big}}_j$). Since $R_j$ is always the same within that set of questions, it does not tell us how reward points should be distributed between the competing users. Instead, the reward distribution must be determined by comparing only the accuracy of their forecasts.  With $R_j$ now fixed, users should be incentivized to provide the most accurate forecasts that they can.
If the reward received by user $i$ on question $j$ is proportional to $s_{i,j}$ then, for this subset of questions, the reward distribution rule is a strictly proper forecast scoring rule. Therefore, for the set of questions with constant $R_j$, I propose to give user $i$ a reward proportional to $(s^{\text{Big}}_j - s_{i,j})$ on each question. The negative sign on $s_{i,j}$ just ensures that a smaller surprisal corresponds to a larger \emph{positive} reward. When $s_{i,j} = s_j^\text{Big}$, user $i$ gets zero reward. 

Combining the first and second steps above results in the following reward distribution rule:
\begin{align}
r_{i,j} &{} = R_j \left(s^{\text{Big}}_j - s_{i,j} \right) \nonumber \\
&{} =  R_j \left( \langle s_j \rangle + c \Delta s_j - s_{i,j} \right) \, . \label{e.rewarddist}
\end{align}
Namely, user $i$ receives a reward on question $j$ proportional to $R_j$ for all fixed values of $s^{\text{Big}}_j - s_{i,j}$, and to $s^{\text{Big}}_j - s_{i,j}$ for all fixed values of $R_j$. The value of $c$ determines how small user $i$'s surprisal $s_{i,j}$ on question $j$ must be relative to that of their peers in order to receive a positive reward.    

Equation \ref{e.rewarddist} can be written compactly as
\begin{equation}
r_{i,j} = \begin{cases} 
	R_j \ln \parz{\frac{p_{i,j}}{p_c}} \, & \text{if} \, q_j = +1 \\
	R_j \ln \parz{\frac{1-p_{i,j}}{p_c}} \, & \text{if} \, q_j = -1 \\
	0 \, & \text{if} \, q_j = 0
	\end{cases} \, , \label{e.finalreward}
\end{equation}
where 
\begin{equation}
p_c = e^{-\langle s_j \rangle - c \Delta s_j } \, ,
\end{equation}
and where I have used equation \ref{e.surprisaldef}.
For the sake of providing numerical illustrations later on, I will always use $c =1$ for the rest of this paper, although it can in principle be re-tuned later to achieve an optimal algorithm. 

The reward distribution algorithm in equation~\ref{e.finalreward} only deals with those users who make predictions. However, as in the preliminary examples from section \ref{single}, 
the users who propose the initial questions to begin with should also receives reward points based on the spread $\Delta s_j$ and the consensus $|V_j|$. I propose to simply make this reward equal to the maximum of the $r_{i,j}$ in equation~\ref{e.finalreward}:
\begin{equation}
\label{e.questionaward}
r_{\text{questioner},j} =
\text{max} \parz{r_{i,j}} \, .
\end{equation}
The questioner gets a reward equal to that of the most accurate forecast. 

One can check that this algorithm reproduces the main features of the narrative in section \ref{ex_quaerum}. 
In scenario \#1 of the tornado example in figure \ref{f.scenarios1th3}, $\left \langle s_j \right \rangle \approx 1.2$, $\Delta s_j \approx 2.3$ and $p_c=0.1$. So for user \#1 who had the smaller surprisal, $r_{1,j} \approx 1.1 \times \ln{\left (9 \right)} \approx 2.4$.  For user \#2 who had the larger surprisal, $r_{2,j} \approx 1.1 \times \ln{\left( 1 \right)}=0$.  In scenario \#2, the reward sizes are the same but reversed for the two users. 
In scenario \#3, $\left|V_j\right|=0$ so $R_j=\left|\Delta s_j V_j \right| = 0$. In scenario \#4, $\Delta s_j = 0$ so  $R_j=\left|\Delta s_j V_j \right| = 0$. So, in both scenario \#3 and scenario \#4, zero reward points get distributed. User \#0, who proposed the initial question in both scenarios \#1 and \#2, gets a large award $r_{0,j}\approx2.4$ for ushering the question successfully from start to finish, see equation \ref{e.questionaward}. In scenarios \#3 and \#4, user \#0 receives zero reward points. A similar check can be applied to scenarios \#5 and \#6, which I will leave to the reader.

Over many questions, a user’s reward score is simply the sum of all rewards for each question,
\begin{equation}
\label{e.totalr}
r_i = \sum_j^{n_i} r_{i,j} \, ,
\end{equation}
where $n_i$ is the total number of predictions ever made by user $i$. The total number of reward points (aside from those given to the questioner) handed out on a single \emph{question} is
\begin{equation}
r_{j,\text{total}} = \sum_i^{N_j} r_{i,j} = c N_j R_j \Delta s_j = c N_j \Delta s_j^2 | V_j | \, ,
\end{equation}
as can be easily checked. 

Recall from equation~\ref{e.surprisaldef} that, strictly speaking, the forecast scoring rule does not necessarily measure the accuracy of forecasting the true outcomes, but rather what the majority \emph{believes} are the outcomes, as measured by the $q_j$ from equation \ref{e.outcome_def}. A low surprisal, in this case, is only a good measure of genuine forecasting skill if $q_j$ is a reliable proxy for the truth of actual question outcomes. A natural worry is that users who understand this will make forecasts based on what they believe is the crowd consensus rather than on their own careful and independent analyses. But if many users start to follow the group, then the predictions will cluster together, $\Delta s_j$ will approach zero, and the reward scale $R_j$ will approach zero. Therefore, a group cannot collect significant reward points on questions where this or similar group-think trends starts to take hold. In order for anyone in a group of users to receive large rewards, they must make predictions that are accurate, but that also deviate from each other by a wide margin. To get a large reward distribution, the group must cultivate independence among its members. 

 Furthermore, questions that never influence users to change their minds will tend to produce a $|V_j|$ close to zero. Again, the award scale $R_j$ will be close to zero in this case so few users can gain any significant reward points on these questions. Consider again our discussion of scenario \#3 in section~\ref{single} in light of the above system.  
  
It is easy to imagine other ways that users might attempt to game this reward system in ways that defeat the goals of section~\ref{s.predictions}, and it will be important to establish mechanisms to counter them. I will have more comments about this in section~\ref{s.future}, but
for now I will just mention one other possible strategy for gaming the system. Consider a user who has no interest in gaining large reward points. They could choose to give \emph{intentionally poor} forecasts for some questions (perhaps at the behest of other users) as a way to inflate $\Delta s_j$ and thus the value of $R_j$. For example, a user could be retired from research and no longer interested in maintaining a high prediction score. This person could begin making intentionally poor predictions at the request of a group of friends as a way to artificially inflate the apparent predictive power of that group. However, with the reward algorithm in equation \ref{e.finalreward} a user who does this will end up with a large \emph{negative} award. This is basically a reward penalty, and if they do it regularly they will accumulate a very large negative reward via equation~\ref{e.totalr}. Notice that a negative reward is actually quite difficult to maintain over many questions since the user needs to make forecasts that are systematically wrong and opposed to the crowd. Therefore, the reward system provides its own mechanism for mitigating the type of system-gaming described above -- the presence of users with an accumulated large negative score can be used as a clear warning that it is taking place. If a question involves significantly many users who carry negative scores from past questions, then this provides a red flag for possible manipulation.

The possibility to receive negative scores might seem troubling at first, but note that the reward system is actually net positive. 
A good-faith but below average forecaster will generally gain positive rewards over many question despite having a relatively poor forecasting record. Examples that we will show in section~\ref{simulations} will illustrate this. Also note that none of the users in scenarios \#1-\#6 from section \ref{ex_quaerum} receive negative rewards, even when they ``lose.''
Allowing for negative reward points dissuades the type of systematic manipulations described above, but does not punish earnest forecasting. The goal is that any good-faith participant will receive net positive reward points for their participation in the system.

\section{Scientific metadata}
\label{s.metadata}

The data for $r_{i,j}$, $s_{i,j}$, etc, accumulated over time, will be especially valuable if users can refer to specific publications, theories, or hypotheses when they formulate forecasts or questions and link to them 
within the \textit{Ex Quaerum} system. Textual information included to explain the rationale behind questions and forecasts would be valuable for placing these numbers in context. Questions and forecasts that are related to one another can be connected and catalogued according to common themes (see again the orange and blue arrows in figure~\ref{f.many}).

Each expert will have their own summary page listing data such as their total reward points $r_i$ from equation \ref{e.totalr} and the total number of questions they have answered, with links to those questions. 
There may also be pages summarizing data for individual questions. Each question page will include data like the average $\left \langle s_j \right \rangle$ and the value of $R_j$. (From here forward, I will drop the question index $j$ on $R_j$ unless it is necessary.)
Over time, there will be an accumulation of data such as average surprisal scores, typical $R$, and other metrics based on the combination of predictive spread, predictive accuracy, and consensus. 

\section{Simulations}
\label{simulations}

With a specific scoring and reward algorithms now set up, it will be easier to step through scenarios that 
illustrate how I intend for the system to function in practice.  
I will spend much of the rest of this paper describing more  scenarios like those in section \ref{ex_quaerum} to help further build intuition, but now using actual numbers and referring back to the equations of section \ref{s.algorithm}.
To maintain continuity with the labeling in section \ref{ex_quaerum}, I will start here with a ``scenario \#7.''

\subsection{Scenario \#7}
\label{sim1}

\begin{table*}[]
\centering
\begin{tabular}{lllllllll}
\cellcolor[HTML]{C6E0B4}User 1 & \cellcolor[HTML]{BDD7EE}Predictions & \cellcolor[HTML]{F8CBAD}Validate? & \cellcolor[HTML]{FFE699}surprisal 1 &  & \cellcolor[HTML]{C6E0B4}User 2 & \cellcolor[HTML]{BDD7EE}Predictions & \cellcolor[HTML]{F8CBAD}Validate? & \cellcolor[HTML]{FFE699}surprisal 2 \\
question 1                     & 0.9                                 & 1                                 & 0.10536052                          &  & question 1                     & 0.7                                 & 1                                 & 0.35667494                          \\
question 2                     & 0.07                                & -1                                & 0.07257069                          &  & question 2                     & 0.4                                 & -1                                & 0.51082562                          \\
question 3                     & 0.1                                 & -1                                & 0.10536052                          &  & question 3                     & 0.3                                 & -1                                & 0.35667494                          \\
question 4                     & 0.96                                & 1                                 & 0.04082199                          &  & question 4                     & 0.88                                & 1                                 & 0.12783337                          \\
question 5                     & 0.99                                & 1                                 & 0.01005034                          &  & question 5                     & 0.99                                & 1                                 & 0.01005034                          \\
                               &                                     &                                   &                                     &  &                                &                                     &                                   &                                     \\
\cellcolor[HTML]{C6E0B4}User 3 & \cellcolor[HTML]{BDD7EE}Predictions & \cellcolor[HTML]{F8CBAD}Validate? & \cellcolor[HTML]{FFE699}surprisal 3 &  & \cellcolor[HTML]{C6E0B4}User 4 & \cellcolor[HTML]{BDD7EE}Predictions & \cellcolor[HTML]{F8CBAD}Validate? & \cellcolor[HTML]{FFE699}surprisal 4 \\
question 1                     & 0.17                                & 1                                 & 1.77195684                          &  & question 1                     & 0.2                                 & 1                                 & 1.60943791                          \\
question 2                     & 0.8                                 & -1                                & 1.60943791                          &  & question 2                     & 0.95                                & 1                                 & 2.99573227                          \\
question 3                     & 0.4                                 & -1                                & 0.51082562                          &  & question 3                     & 0.7                                 & 1                                 & 1.2039728                           \\
question 4                     & 0.8                                 & 1                                 & 0.22314355                          &  & question 4                     & 0.65                                & 1                                 & 0.43078292                          \\
question 5                     & 0.95                                & 1                                 & 0.05129329                          &  & question 5                     & 0.85                                & 1                                 & 0.16251893                          \\
                               &                                     &                                   &                                     &  &                                &                                     &                                   &                                     \\
\cellcolor[HTML]{C6E0B4}User 5 & \cellcolor[HTML]{BDD7EE}Predictions & \cellcolor[HTML]{F8CBAD}Validate? & \cellcolor[HTML]{FFE699}surprisal 5 &  &                                &                                     &                                   &                                     \\
question 1                     & 0.19                                & 1                                 & 1.66073121                          &  &                                &                                     &                                   &                                     \\
question 2                     & 0.95                                & 1                                 & 2.99573227                          &  &                                &                                     &                                   &                                     \\
question 3                     & 0.6                                 & -1                                & 0.91629073                          &  &                                &                                     &                                   &                                     \\
question 4                     & 0.8                                 & 1                                 & 0.22314355                          &  &                                &                                     &                                   &                                     \\
question 5                     & 0.9                                 & 1                                 & 0.10536052                          &  &                                &                                     &                                   &                                    
\end{tabular}
    \caption{Scenario \#7 is a simulation of five question-prediction-resolution cycles with five users. The rewards each user accumulated from 
    this sequence of questions are shown in figure \ref{f.sim1}. Under the ``validate'' column, $1$ means the user claimed ``yes'' as the outcome and $-1$ means the user claimed ``no.''}
\label{table1}
\end{table*}
%
\begin{figure*}
\centering
\includegraphics[scale=0.55]{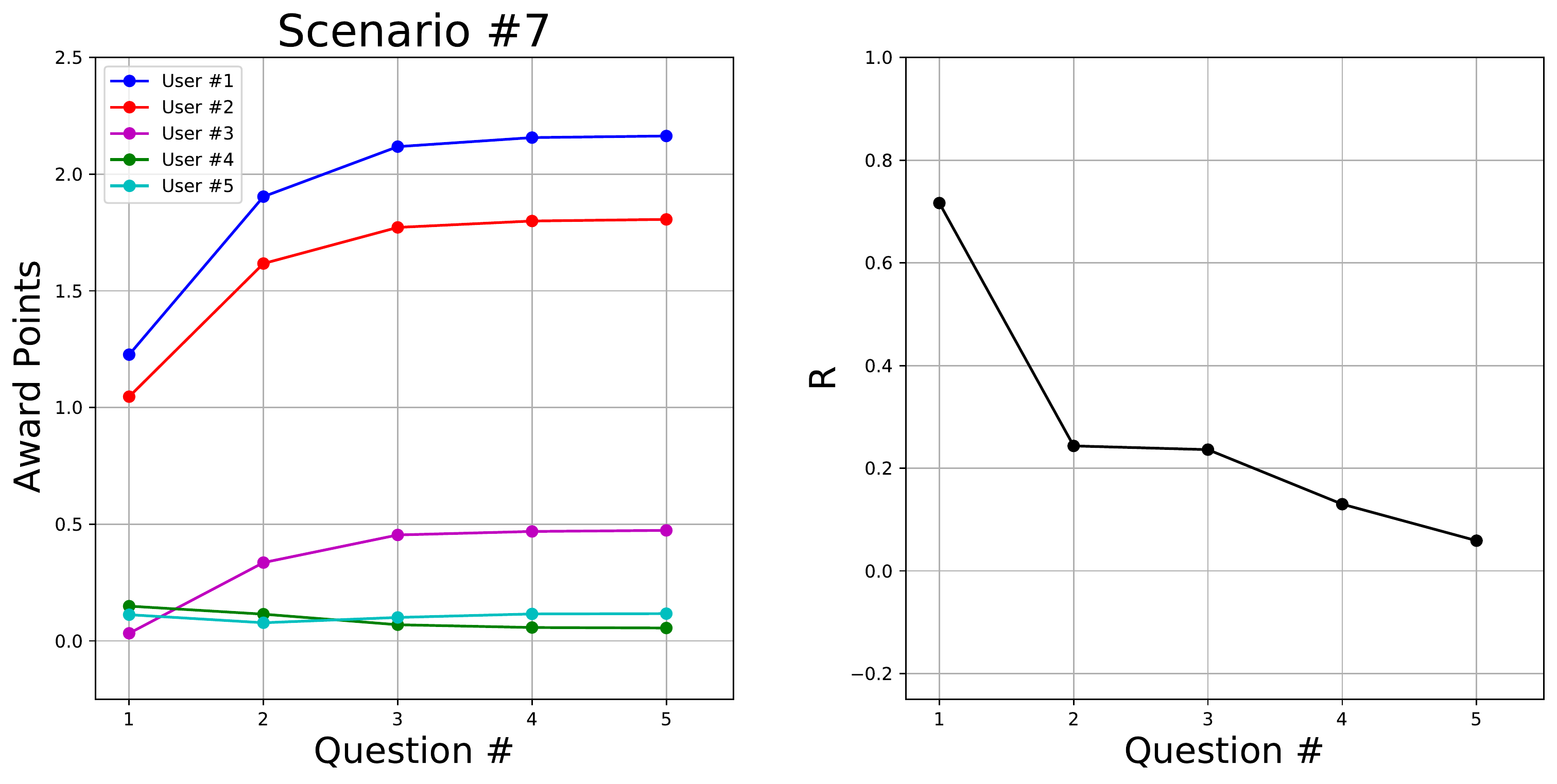}
\caption{A summary of rewards based on the results in table~\ref{table1}. The left hand plot shows the \emph{\underline{total}} number of 
reward points accumulated by each user over the course of the five questions, while the right hand plot is the reward scale $R$, defined in equation~\ref{e.reward_scale}, for each question. Note from right hand plot that the potential reward payout drops as the users convergent in agreement around the Stockholm model.}
\label{f.sim1}
\end{figure*}
\begin{figure*}
\centering
\includegraphics[scale=0.55]{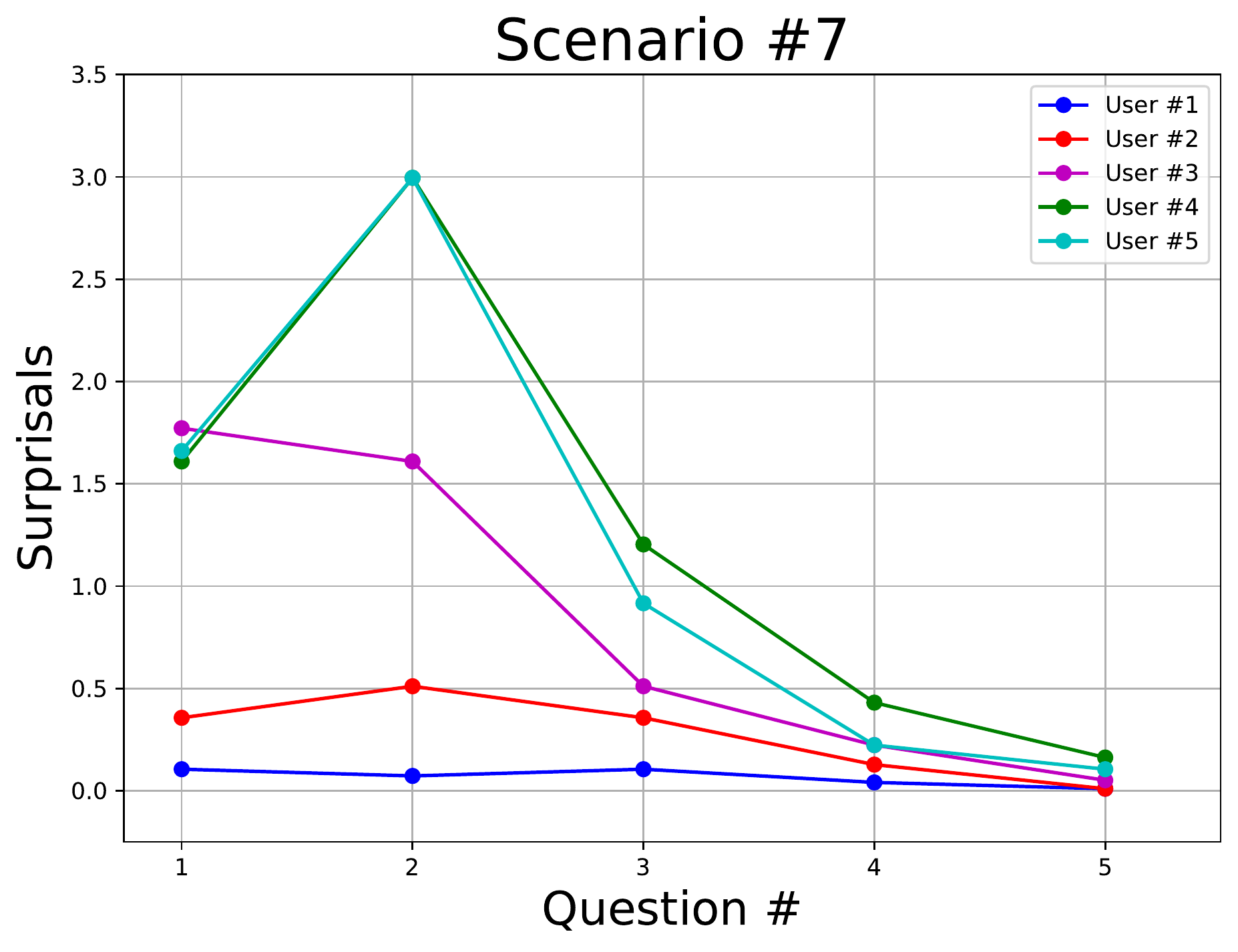}
\caption{The distributions of surprisals corresponding to each user in the question sequence of Table~\ref{table1}.}
\label{f.sur1}
\end{figure*}

Let there now be five expert users all submitting predictions over a sequence of five questions. To keep a vivid narrative, we might imagine that this still involves the climate scientists of section \ref{single}, some of whom are adherents of the Stockholm model and some of whom prefer the Chicago model. Still other users might be undecided about which model is more reliable. Table~\ref{table1} displays what a sequence of prediction outcomes might look like. I have generated these numbers mostly by hand just to illustrate the basic mechanics of the scoring and reward distribution. The total number of reward points attained by each user is plotted in figure~\ref{f.sim1} and the corresponding surprisals are shown in figure~\ref{f.sur1}.

The following is a plausible scenario \#7 narrative for how the different expert users might have reached their forecasts: 

From the outset, users \#1 and \#2 are strong adherents of the newer Stockholm model. User \#3 is not yet an adherent, but is somewhat open to being convinced. Users \#4 and \#5 are strong adherents of the more traditional Chicago climate model. Thus, on the first question users \#1 and \#2 lean toward a 
``yes'' answer, but users \#3, \#4, and \#5 lean toward ``no.''
This is reflected in their forecast probabilities listed in the ``question 1'' rows of Table~\ref{table1} under the ``Prediction'' columns. All five expert users agree, as indicated by the ``$1$'' under the ``Validate?'' columns, that the outcome was ``yes.'' The result is a rather large surprisal for the Chicago model adherents, and large reward (small surprisal) for the Stockholm  model adherents. 

 After question \#1, users \#3, \#4, and \#5 are nevertheless unswayed from their confidence in the Chicago model relative to the Stockholm model. They feel that the 
arguments in favor of the Chicago model are too powerful to be abandoned so easily, and that the outcome of the first question is likely an unlucky fluke. Thus, for question 2 there is again a wide range of different predictions. Users \#1 and \#2 give fairly small probabilities for ``yes,''  \#3, \#4, and \#5 believe the outcome will more likely be ``no.'' This time there is poor agreement over how the prediction for question 2 resolved. Users \#4 and \#5 insist that the outcome was indeed ``no'' but  \#1 and \#2 insist it was ``yes.'' User \#3 now breaks the tie and validates the outcome as ``no,'' despite having given a relatively large forecast for ``yes.'' In fact, the outcome of question 2 begins to sway user \#3 more toward supporting the Stockholm model and rejecting the Chicago model. 

The absolute value of the mean validation (the ``consensus'') for question 2 is only $0.2$ because the agreement is only $3$-to-$2$ in favor of ``no.'' Hence, the value of the reward scale $R$ for question 2 is rather small, and fewer award points can be distributed -- see the right-hand plot in \ref{f.sim1} and compare $R$ for question 2 with that of question 1. A question-prediction-resolution cycle like question 2 will generally be considered less successful due to the smaller $R$. 

Given the disappointing lack of consensus in question 2, the users now make an extra effort to formulate question 3 more carefully to avoid future disagreements about outcomes. They only agree to make additional predictions after long discussions about how to ensure that any possible future outcome disputes have been addressed. 

For question 3, users \#1, \#2, and \#3 all give a low probabilities for ``yes'' based on their growing confidence in the Stockholm  model. Users \#4 and \#5 give a moderately high probabilities, again based on the Chicago model. The outcome of question 3 again comes out in favor of the Stockholm  model, now with all users convinced that the outcome was ``no,'' except for user \#4 who remains reluctant to validate the outcome in a way that favors the Stockholm model. The absolute value of the mean validation is $0.6$ for question 3. 

With the first three questions generally favoring the Stockholm model, all users now begin to defer to it when making predictions. The last predictions, for questions \#4 and \#5, are accurate and begin to cluster together. They are validated with unanimous consensus for ``yes.'' Since all users' predictions on questions \#4 and \#5 are similar, the $R$ values are very small, and only a very small number of award points are available to be distributed. All users gain only a very tiny boost to their total number of reward points, now not visible on the graph. 

All five users have earned a positive number of reward points at the end of the five-question sequence, though users \#4 and \#5 have only a very small gain since their forecasts were consistently the least accurate. Users \#1 and \#2 have the highest overall reward, while user \#3 has a respectable gain. 

Although the number of users and questions is still small here, comparing the Stockholm  and Chicago models shows the beginnings of a trend, even to someone with no knowledge of the science underlying the competing models. Most importantly, an external observer does not need to rely on specific arguments from adherents of either the Stockholm or Chicago model to see what the trend is. 
The distribution of surprisals in figure~\ref{f.sur1} visibly clusters around small values at the same time that the group of users turns toward the Stockholm model. Notice that the Stockholm model skeptics have played an important role here, despite their low scores, because outside observers can see that the trend involves Chicago model adherents conceding to Stockholm model outcomes. They have provided the checks on any biases or blind spots that the Stockholm adherents might have.

Retracing the reasoning of Chicago and Stockholm adherents is further helped if the expert users have recorded explanations of their predictions and included links to journal articles at each step in the process.

\subsection{Scenario \#8}
\label{sim2}

The sequence of questions in figure \ref{f.sim1} involved a reasonably wide range of 
predictions as indicated by the range of surprisal scores, at least at its outset, so there was a high potential, as measured by the size of $R$, for users to gain reward points. 
This was by design. A wide range of predictions implies a wide range of competing viewpoints. The members of the group thus act as checks on one another when they assess outcomes.  

\begin{figure*}
\centering
\includegraphics[scale=0.55]{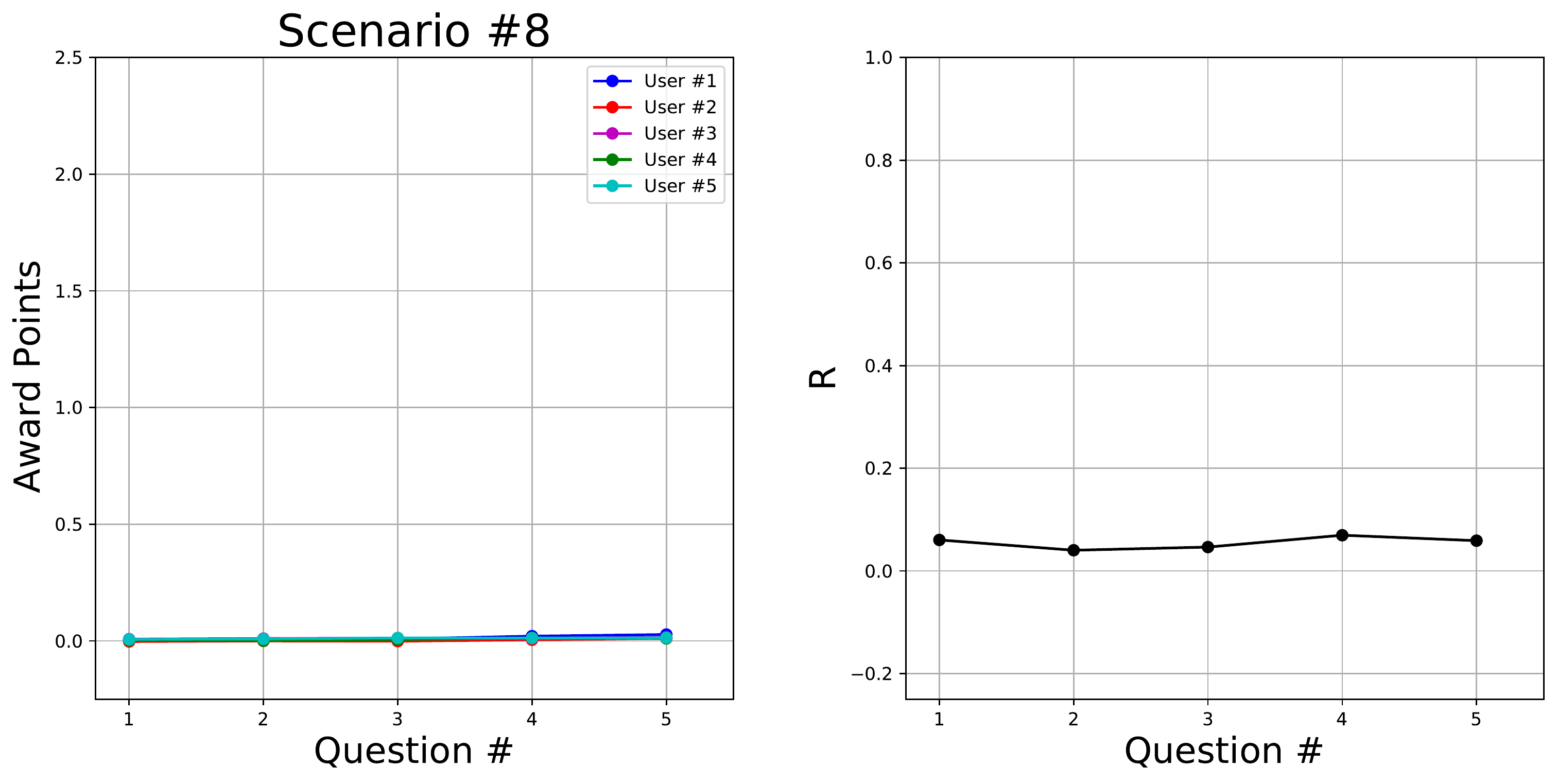}
\caption{Now the reward points that are distributed to users \#1 through 
\#5 based on the scenario \#8 results in \ref{table2}. The scales of plot axes are the same as in~\ref{f.sim1} to emphasize the smallness of the rewards as compared to scenario \#1.}
\label{f.sim2}
\end{figure*}

\begin{figure*}
\centering
\includegraphics[scale=0.55]{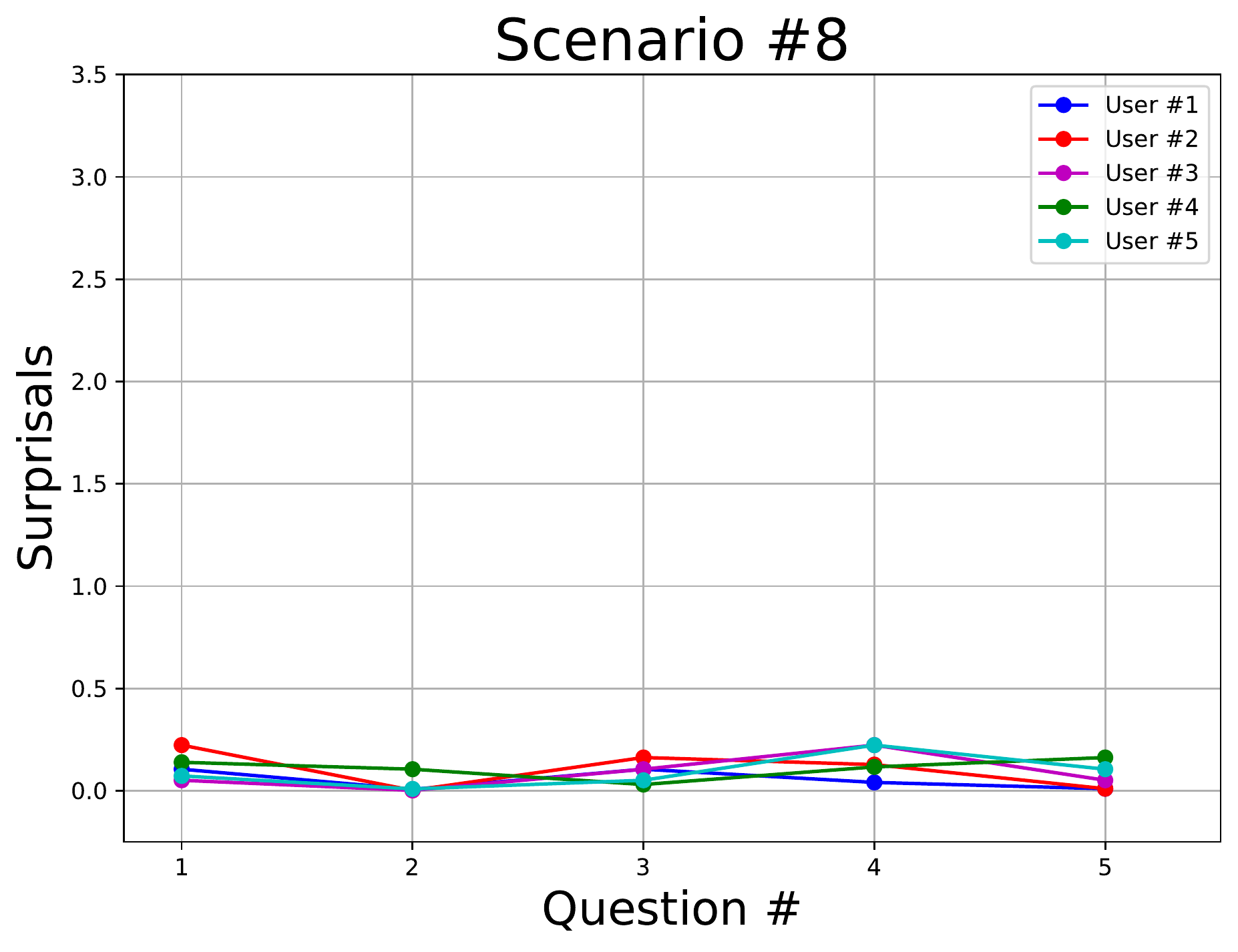}
\caption{The distributions of surprisals associated with each user in the scenario \#8 question sequence of Table~\ref{table2}. There has been no discernible evolution in prediction accuracy over the five-question sequence.}
\label{f.sur2}
\end{figure*}

It is instructive to contrast this with what would happen in a similar situation but where all users are all in agreement, and the outcomes confirm the consensus expectation. I will label this as scenario \#8. Since all users make similar predictions, the group does not have any  check on the consensus view. So, in a scenario like this the chances for earning a large award are limited by design. 
An example is in table~\ref{table2} and in figures~\ref{f.sim2} and~\ref{f.sur2}. Here, all five users are in strong agreement about the likelihood of each outcome, and in all five questions they confirm that the most expected outcome is indeed what was observed. Some reward points get distributed, but the amounts are quite small compared to scenario \#7, and I have emphasized this in figure \ref{f.sim2} by keeping the vertical axes the same as in figure \ref{f.sim1}. Figure \ref{f.sur2} implies to any onlookers that there has been very little movement in the consensus view as measured by surprisal. 
\begin{table*}[]
\begin{tabular}{lllllllll}
\cellcolor[HTML]{C6E0B4}User 1 & \cellcolor[HTML]{BDD7EE}Predictions & \cellcolor[HTML]{F8CBAD}Validate? & \cellcolor[HTML]{FFE699}surprisal 1 &  & \cellcolor[HTML]{C6E0B4}User 2 & \cellcolor[HTML]{BDD7EE}Predictions & \cellcolor[HTML]{F8CBAD}Validate? & \cellcolor[HTML]{FFE699}surprisal 2 \\
question 1                     & 0.9                                 & 1                                 & 0.10536052                          &  & question 1                     & 0.8                                 & 1                                 & 0.22314355                          \\
question 2                     & 0.007                               & -1                                & 0.00702461                          &  & question 2                     & 0.002                               & -1                                & 0.002002                            \\
question 3                     & 0.1                                 & -1                                & 0.10536052                          &  & question 3                     & 0.15                                & -1                                & 0.16251893                          \\
question 4                     & 0.96                                & 1                                 & 0.04082199                          &  & question 4                     & 0.88                                & 1                                 & 0.12783337                          \\
question 5                     & 0.99                                & 1                                 & 0.01005034                          &  & question 5                     & 0.99                                & 1                                 & 0.01005034                          \\
                               &                                     &                                   &                                     &  &                                &                                     &                                   &                                     \\
\cellcolor[HTML]{C6E0B4}User 3 & \cellcolor[HTML]{BDD7EE}Predictions & \cellcolor[HTML]{F8CBAD}Validate? & \cellcolor[HTML]{FFE699}surprisal 3 &  & \cellcolor[HTML]{C6E0B4}User 4 & \cellcolor[HTML]{BDD7EE}Predictions & \cellcolor[HTML]{F8CBAD}Validate? & \cellcolor[HTML]{FFE699}surprisal 4 \\
question 1                     & 0.95                                & 1                                 & 0.05129329                          &  & question 1                     & 0.87                                & 1                                 & 0.13926207                          \\
question 2                     & 0.002                               & -1                                & 0.002002                            &  & question 2                     & 0.1                                 & -1                                & 0.10536052                          \\
question 3                     & 0.1                                 & -1                                & 0.10536052                          &  & question 3                     & 0.03                                & -1                                & 0.03045921                          \\
question 4                     & 0.8                                 & 1                                 & 0.22314355                          &  & question 4                     & 0.89                                & 1                                 & 0.11653382                          \\
question 5                     & 0.95                                & 1                                 & 0.05129329                          &  & question 5                     & 0.85                                & 1                                 & 0.16251893                          \\
                               &                                     &                                   &                                     &  &                                &                                     &                                   &                                     \\
\cellcolor[HTML]{C6E0B4}User 5 & \cellcolor[HTML]{BDD7EE}Predictions & \cellcolor[HTML]{F8CBAD}Validate? & \cellcolor[HTML]{FFE699}surprisal 5 &  &                                &                                     &                                   &                                     \\
question 1                     & 0.93                                & 1                                 & 0.07257069                          &  &                                &                                     &                                   &                                     \\
question 2                     & 0.009                               & -1                                & 0.00904074                          &  &                                &                                     &                                   &                                     \\
question 3                     & 0.05                                & -1                                & 0.05129329                          &  &                                &                                     &                                   &                                     \\
question 4                     & 0.8                                 & 1                                 & 0.22314355                          &  &                                &                                     &                                   &                                     \\
question 5                     & 0.9                                 & 1                                 & 0.10536052                          &  &                                &                                     &                                   &                                    
\end{tabular}
\caption{The scenario \#8 question-prediction-resolution sequence is similar to Table~\ref{table1}, but now all users are in close agreement with regard to both their predictions and their validations. The corresponding award distributions are shown in figure \ref{f.sim2} and the surprisals are in figure~\ref{f.sur2}.}
\label{table2}
\end{table*}

%

\subsection{Scenario \#9}

The previous example might leave the impression that 
questions on which all users agree with a high probability in one direction always lead to a low reward distribution. This is not the case, which I can demonstrate with a slight modification to scenario \#8. 

Notice from table~\ref{table2} of scenario \#8 that all users gave a very low probability for ``yes'' on question 2. All users apart from user \#4 gave a probability of less than $1.0\%$. When the outcome of the question is ``no,'' as in scenario \#8, very few reward points are distributed. Indeed, one might suspect that user \#4 was overly conservative with their forecast. 

But if the outcome of question 2 is ``yes,'' the spread in surprisals is much larger. I will call this scenario \#9. It is shown in  table~\ref{table3} and it is identical to table~\ref{table2} except that all users agree that, despite expectations, the answer to question 2 turned out to be ``yes.'' User \#4 gave only a $1.0\%$ chance of a ``yes'' outcome for question 2, but they nevertheless receive a very large reward because other users gave even smaller chances. User \#4 has been rewarded for their cautious forecast in scenario \#9. Averaged over many questions with the same $R$, the smallest surprisal goes to the users whose forecasts most accurately represent the true probabilities. 

This last scenario illustrates that large reward points are possible whenever there is an unexpected outcome, even if all users were originally in agreement about the most likely result.

In the three scenarios above, all the users have earned a positive number of reward points at the end of the five question sequence. On top of the points shown in the figures, which they have earned from predictions, they might also have earned award points according to equation~\ref{e.questionaward} by proposing successful questions. 

Scenarios \#7-\#9 illustrate the strategies an expert user should follow if they wish to accumulate a large amount of reward points over many questions. A user must: i.) find questions with a wide spread of different predictions by seeking out topics that bring disagreements and/or uncertainties to the forefront, ii.) find competing users who are nevertheless willing to concede to a consensus given clear evidence, iii.) find questions that are formulated clearly enough that an eventual consensus is likely, iv.) regularly cast at least reasonably well calibrated forecasts relative to their peers and/or, v.) propose successful questions. (Again, "successful'' means a question with a large $R$. It refers to a question that begins with a wide range of different predictions and a wide range in surprisals, but ends with a high degree of consensus.)

\begin{table*}[]
\begin{tabular}{lllllllll}
\cellcolor[HTML]{C6E0B4}User 1 & \cellcolor[HTML]{BDD7EE}Predictions & \cellcolor[HTML]{F8CBAD}Validate? & \cellcolor[HTML]{FFE699}surprisal 1 &  & \cellcolor[HTML]{C6E0B4}User 2 & \cellcolor[HTML]{BDD7EE}Predictions & \cellcolor[HTML]{F8CBAD}Validate? & \cellcolor[HTML]{FFE699}surprisal 2 \\
question 1                     & 0.9                                 & 1                                 & 0.10536052                          &  & question 1                     & 0.8                                 & 1                                 & 0.22314355                          \\
question 2                     & 0.007                               & -1                                & 0.00702461                          &  & question 2                     & 0.002                               & -1                                & 0.002002                            \\
question 3                     & 0.1                                 & -1                                & 0.10536052                          &  & question 3                     & 0.15                                & -1                                & 0.16251893                          \\
question 4                     & 0.96                                & 1                                 & 0.04082199                          &  & question 4                     & 0.88                                & 1                                 & 0.12783337                          \\
question 5                     & 0.99                                & 1                                 & 0.01005034                          &  & question 5                     & 0.99                                & 1                                 & 0.01005034                          \\
                               &                                     &                                   &                                     &  &                                &                                     &                                   &                                     \\
\cellcolor[HTML]{C6E0B4}User 3 & \cellcolor[HTML]{BDD7EE}Predictions & \cellcolor[HTML]{F8CBAD}Validate? & \cellcolor[HTML]{FFE699}surprisal 3 &  & \cellcolor[HTML]{C6E0B4}User 4 & \cellcolor[HTML]{BDD7EE}Predictions & \cellcolor[HTML]{F8CBAD}Validate? & \cellcolor[HTML]{FFE699}surprisal 4 \\
question 1                     & 0.95                                & 1                                 & 0.05129329                          &  & question 1                     & 0.87                                & 1                                 & 0.13926207                          \\
question 2                     & 0.002                               & -1                                & 0.002002                            &  & question 2                     & 0.1                                 & -1                                & 0.10536052                          \\
question 3                     & 0.1                                 & -1                                & 0.10536052                          &  & question 3                     & 0.03                                & -1                                & 0.03045921                          \\
question 4                     & 0.8                                 & 1                                 & 0.22314355                          &  & question 4                     & 0.89                                & 1                                 & 0.11653382                          \\
question 5                     & 0.95                                & 1                                 & 0.05129329                          &  & question 5                     & 0.85                                & 1                                 & 0.16251893                          \\
                               &                                     &                                   &                                     &  &                                &                                     &                                   &                                     \\
\cellcolor[HTML]{C6E0B4}User 5 & \cellcolor[HTML]{BDD7EE}Predictions & \cellcolor[HTML]{F8CBAD}Validate? & \cellcolor[HTML]{FFE699}surprisal 5 &  &                                &                                     &                                   &                                     \\
question 1                     & 0.93                                & 1                                 & 0.07257069                          &  &                                &                                     &                                   &                                     \\
question 2                     & 0.009                               & -1                                & 0.00904074                          &  &                                &                                     &                                   &                                     \\
question 3                     & 0.05                                & -1                                & 0.05129329                          &  &                                &                                     &                                   &                                     \\
question 4                     & 0.8                                 & 1                                 & 0.22314355                          &  &                                &                                     &                                   &                                     \\
question 5                     & 0.9                                 & 1                                 & 0.10536052                          &  &                                &                                     &                                   &                                    
\end{tabular}
\caption{A scenario \#9 question-prediction-resolution sequence identical to Table~\ref{table2} but now question 2 has been validated to ``yes'' by all users. The award distributions are shown in figure \ref{f.sim3} and the surprisals are in figure~\ref{f.sur3}.}
\label{table3}
\end{table*}
%
\begin{figure*}
\centering
\includegraphics[scale=0.55]{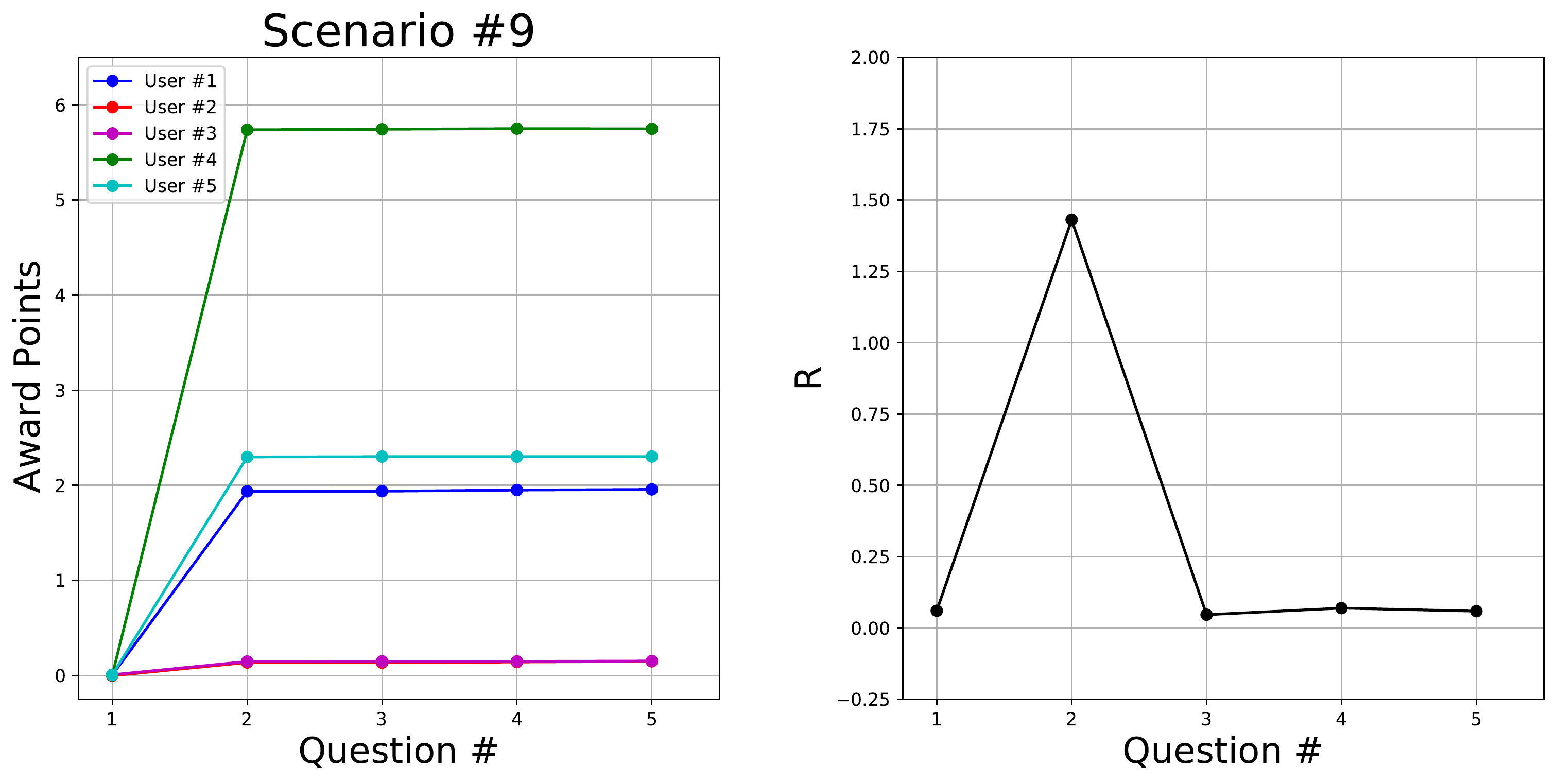}
\caption{Now the reward points that are distributed to users \#1 through 
\#5 based on the scenario \#9 results in \ref{table2}. The vertical axes are expanded relative to figure in~\ref{f.sim1}.}
\label{f.sim3}
\end{figure*}
\begin{figure*}
\centering
\includegraphics[scale=0.55]{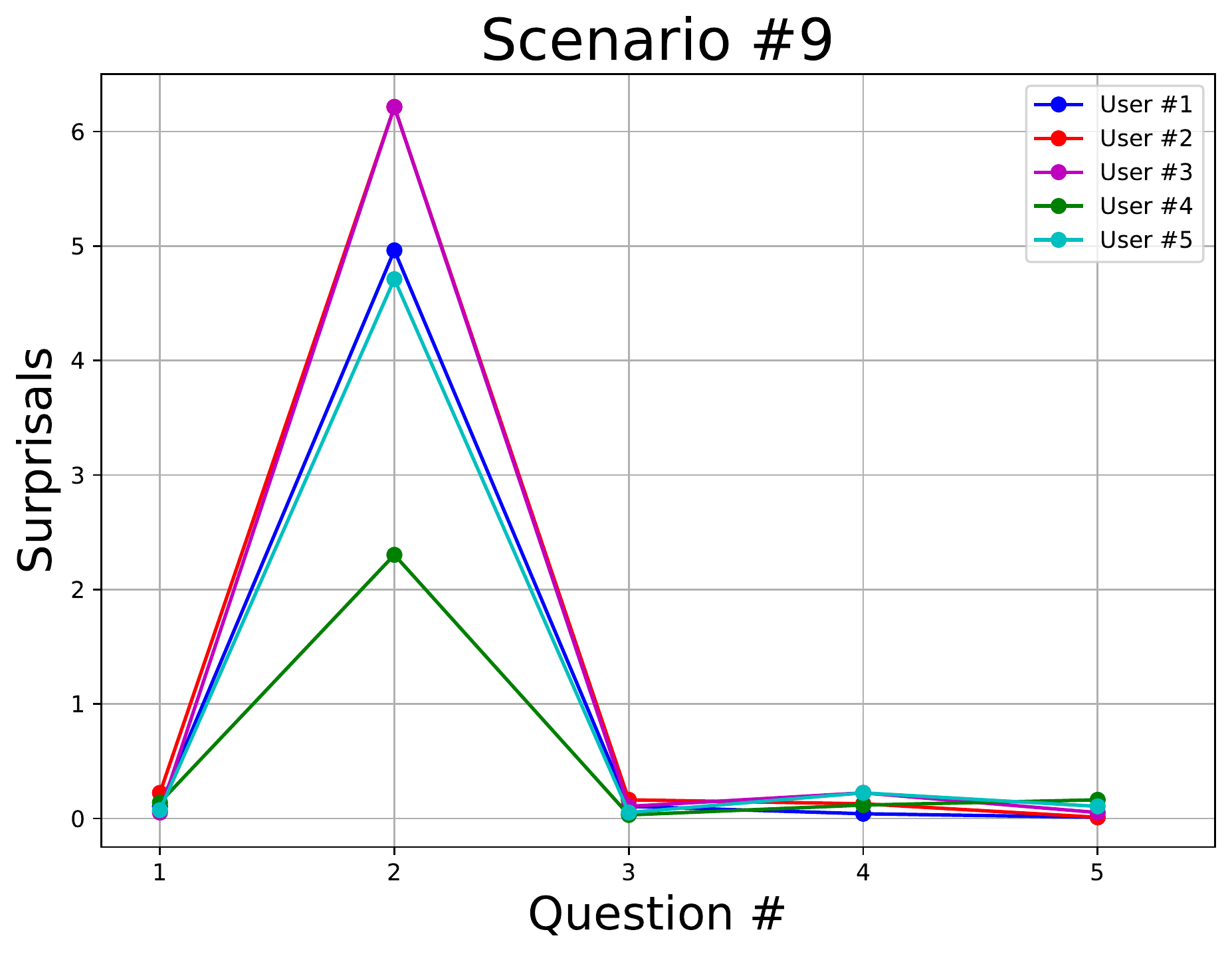}
\caption{The distributions of surprisals associated with each user in the question sequence of Table~\ref{table3}. The vertical axis has been adjusted relative to that of figure~\ref{f.sim1} to accommodate  the lines. The user \#2 and \#3 lines are almost indistinguishable from one another.}
\label{f.sur3}
\end{figure*}
\begin{figure*}
\centering
\begin{tabular}{c}
\includegraphics[scale=0.5]{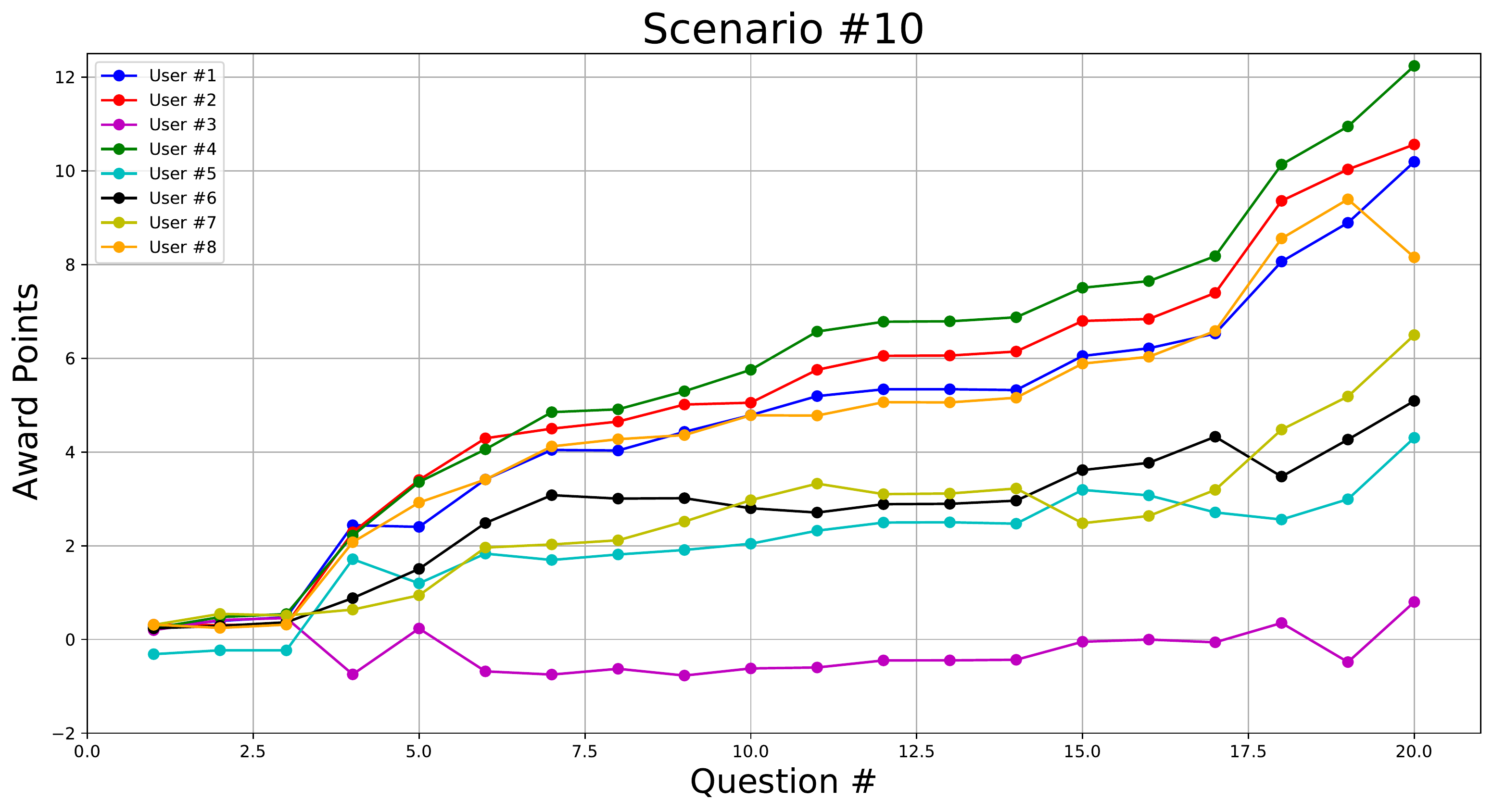} \\
\includegraphics[scale=0.5]{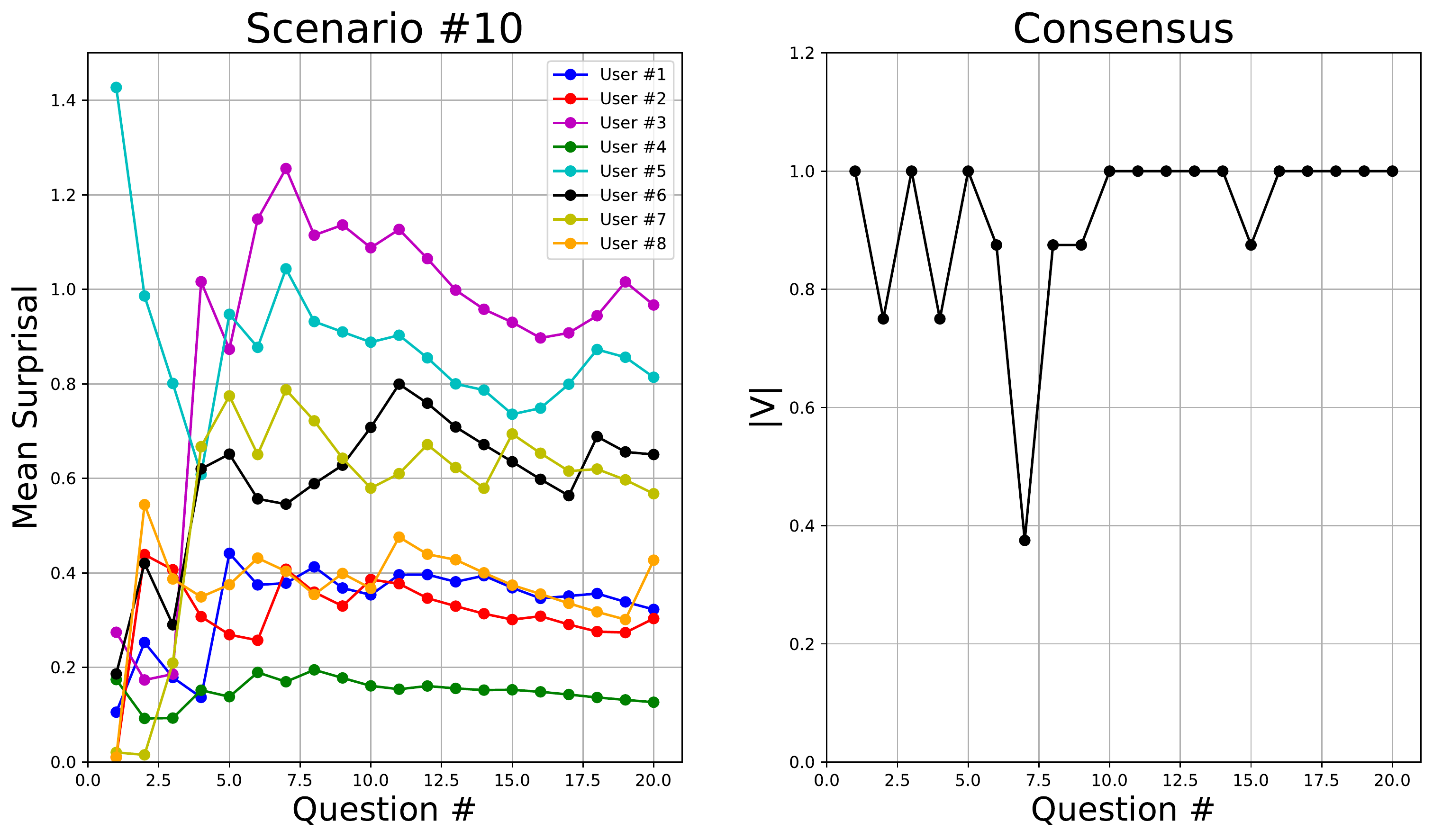}
\end{tabular}
\caption{A simulation involving eight predicting users. The top graph shows the number of award points each accumulates over the course of twenty question-prediction-validation cycles. The lower left graph shows how each user's mean surprisal evolves with each question-prediction-validation cycle. The lower right figure shows the degree of consensus, as measured by the absolute value of equation~\ref{e.meanV}, on each question outcome. A value of $|V| = 0$ means the users were evenly split over what the outcome of the question was. See text for more explanation.}
\label{f.simmany}
\end{figure*}
%
\subsection{Further simulations}
\label{furthersims}

The specific scoring and award recipe I proposed in section~\ref{s.algorithm} is only valuable insofar as it achieves the incentive 
goals set in section~\ref{s.predictions}. It is likely that it will need to be modified, refined, and updated as more is learned about how a system like this works in practice. One way to do this is to run simulations with much larger numbers of hypothetical users and questions in ways that better reflect how actual question-prediction-resolution cycles might play out. 

One more illustrative example is shown in figure~\ref{f.simmany}. 
 Here I have run a small simulation with eight users where I have programmed each to be differently skilled at making forecasts.  The plots show their outcomes on twenty question-resolution-validation cycles. The top plot is each user's accumulated award points on these twenty questions. The bottom left plot shows each user's mean surprisal up to a given question. This plot shows how each user's forecasting skill starts to become evident over the twenty questions. User \#3 is programmed to choose a probability between $0$ and $1$ on each question entirely at random, and this provides a baseline against which to compare all the other users. After many questions, user \#3's surprisal approaches 1.  User \#4 makes very accurate forecasts, with a typical surprisal less than $0.2$. The bottom right figure displays the degree of consensus for each question as measured by the absolute value of the mean validation, equation~\ref{e.meanV}. A table of numbers like those in scenarios \#7-\#9 could also be made here, but it is not instructive and it is now very long, so I have not included it.

\section{Future directions}
\label{s.future}

My main purpose with this document is to explain, in the most concise terms possible, one proposal for simultaneously linking measures of 
scientific progress to factors like predictive power and group consensus in a reward distribution protocol. Of course, many other issues still need to be taken into account before something like this can be transformed into a practical and useful system, and I will comment on some of them here.

\subsection{Limitations of the algorithm}

In the reward distribution algorithm of section~\ref{s.algorithm}, I chose to optimize for simplicity over sophistication.  For example, all forecasts in the examples from sections~\ref{ex_quaerum} and \ref{simulations} are for strictly binary ``yes'' or ``no'' questions. 
Furthermore, timing does not, so far, enter at all into the reward distribution scheme. It makes no difference if a user submits their forecast one year or one week before the validation stage. All that matters in section~\ref{s.algorithm} is that predictions are made sometime before outcomes are known.   

Other prediction competition systems have more sophisticated forecast scoring algorithms. Metaculus and Good Judgment Open, for example, reward earlier forecasts more highly than later ones. They also accommodate questions with a range of possible outcomes rather than simply ``yes'' or ``no'' answers. The basic scoring rule of Metaculus is, in fact, very similar to equation~\ref{e.finalreward} of this paper. The new features of equation~\ref{e.finalreward} (and equations~\ref{e.reward_scale} and~\ref{e.questionaward}) lie in the ways that the spread of competing predictions, the final consensus, and the question success are folded into the award distribution scheme, and this adds significant new layers of complexity. 

In future iterations, it might become necessary to incorporate the more sophisticated features like the range of outcomes and/or time sensitive aspects.  However, it may also turn out that the resulting loss of transparency and simplicity is too costly, and unnecessary given the goals and the nature of the questions involved. These are rather different from what exist in most other types of forecasting competitions. For instance, a typical question in Good Judgment Open might involve, say, the outcome of an election, something that is generally much easier to forecast the night before than a year before. For questions like these, factoring in timing is an important part of assessing forecasting skill. By contrast, a  typical question for \text{Ex Quaerum} might involve the outcome of a trial for a new cancer treatment. Then one generally cares very little if the prediction comes immediately before the trial begins or several months in advance. What is far more important in a question like this is that there is a high participation rate by competing cancer experts and that the predictions are actually made before the trial rather than after. Scoring predictions according to their timing might in this case bring in little valuable information and it might have the downside of dissuading participation. 

The purpose of \textit{Ex Quaerum} is to motivate experts to record their predictive reasoning in a transparent format more than it is to capture all the nuances involved in numerical forecast assessment. Thus, the greatest challenge will be to convince significant numbers of participants to join. It may therefore be preferable to maintain a system that is easy to understand but limited in what it measures than one that is sophisticated but involves intricate systems of formulas. Nevertheless, I plan to continue to explore options for enhancing, modifying, or improving the scoring algorithm as needed in the future.

\subsection{Membership control}

The goals of section~\ref{ex_quaerum} are met when the system is as open as possible. Anyone serious about making competing predictions should be able to take part without difficulty. However, a basic level of membership control and management will likely be necessary in practice. Partly this is because the ability to open and close multiple accounts anonymously would create obvious opportunities to trick the system and undermine the incentive structures. As with any forum wherein many participants submit textual content, some basic code of conduct will be necessary. 
I plan to study the effectiveness of different membership control mechanisms, and how to balance them with openness.

\subsection{Anonymity or pseudo-anonymity}

In the example scenarios I have used throughout this document, users were identified only by numerical IDs. With regard to their activities on individual questions, it is in principle never necessary to reveal a user's identity explicitly. In practice, maintaining some level of anonymity might enhance the value of question-prediction-validation data. The only verifiable information about individual users that is is absolutely necessary is their \emph{total} number of award points. Generally, understanding the \emph{exact} type of data that should be made publicly available, and to what extent it is traceable to the activity of specific users, is another aspect of the system that needs to be studied in future updates. 

\subsection{Simulations and limited trials}

The fundamental assumption necessary for the reward algorithm to be valuable is that, on average, the outcomes $q_j$ from equation \eqref{e.outcome_def} will correspond to an objective underlying truth. If it does, then the reward distribution system guides the group toward correct answers. I conjecture that $q_j$ will correspond on average to the objective underlying truth so long as the typical expert participant act reasonably and in good faith. I further conjecture that, under those conditions, the system will be robust against the influence of both biases, misaligned incentives, and small numbers of outlier participants who operate pathologically. Demonstrating exactly how robust it can be will require simulations of large numbers of competing expert scientists, complete with models of biases. This is the primary next step in establishing proof of principle. The results could show that the system in section~\ref{s.algorithm} needs tweaking. The simulations can also provide guidance on some of the issues relevant to an actual eventual implementation, such as the membership control and anonymity problems discussed above. There are likely many interesting problems in decision and game theory involved.

The above is only a limited set of the issues that need to be tested going forward. This document is meant only to provide the initial outlines of an idea, and there is obviously still much work that is necessary to transform it into a practical and useful tool. 
A final goal of this paper is to attract additional collaborators who share the concerns laid out in sections~\ref{s.problem} and~\ref{s.predictions}. 

\section*{Acknowledgments}

I thank Old Dominion University's 2021-2022 
computer science 420/421 class for many discussions that helped sharpen the ideas that I have laid out here. In particulur, I thank Taylor Brett, Janet Brunelle, Jena Essary, Ashish Kondaka, Nathan Livingston, Aaron Williams, and Craig Woodington. I also thank Amber Boehnlein, Yaohang Li, Lucia Tabacu, Balsa Terzic, William Rogers, and Nobuo Sato for many helpful discussions that have influenced the final version of this draft. 


\bibliographystyle{plainurl}
\bibliography{bibliography}


\end{document}